\documentclass[10pt, journal,letterpaper, onecolumn]{IEEEtran}
\usepackage[utf8]{inputenc} 
\usepackage[T1]{fontenc}
\usepackage{url}
\usepackage{ifthen}
\usepackage[normalem]{ulem}
\usepackage{enumitem}
\usepackage{cite}
\usepackage[final]{changes}
\definechangesauthor[name={Revision 2}, color=blue]{R2}
\usepackage[cmex10]{amsmath} 

\usepackage{bbm}
\def\eps{\varepsilon}
\usepackage[cmex10]{amsmath}
\usepackage{amssymb,amsthm,bbm,mathtools}
\usepackage{mleftright}
\mleftright
\usepackage[dvipsnames]{xcolor}
\usepackage[normalem]{ulem}
\usepackage{graphicx}
\usepackage{booktabs, hyperref}
\usepackage{algorithm}
\usepackage{algorithmicx}
\usepackage{algpseudocode}

\def\bR{{\mathbb R}}

\def\bB{{\mathbb B}}

\def\bD{{\mathbb D}}
\def\calY{{\mathcal Y}}

\def\argmin{\mathop{\rm arg\, min}}

\newtheorem{theorem}{Theorem}
\newtheorem{assumption}{Assumption}
\newtheorem{proposition}{Proposition}
\newtheorem{corollary}{Corollary}
\newtheorem{lemma}{Lemma}
\newtheorem{definition}{Definition}

\begin{document}
\title{{Staircase Mechanisms: Optimal Additive Noise for Differential Privacy in Any Dimension}} 


\author{%
  \IEEEauthorblockN{James Melbourne$^*$, Mario Diaz$^\dagger$, and Shahab Asoodeh$^\ddagger$}\\
  \IEEEauthorblockA{$^*$CIMAT, Mexico, $^\dagger$IIMAS, Mexico, $^\ddagger$McMaster University, Canada}
}

\maketitle

\begin{abstract}
We continue a line of work on the optimality of the staircase mechanisms in differential privacy. For arbitrary dimension $n$ and any norm-monotone cost function, we prove that a staircase mechanism minimizes expected cost among all query-independent additive mechanisms satisfying the same privacy guarantees, generalizing and unifying Geng and Viswanath \cite{geng2015optimal} and Kulesza, Suresh, and Wang's \cite{Staircase_General} recent extension of Geng, Kairouz, Oh, and Viswanath~\cite{geng2015staircase}. Our proof leverages convex rearrangement theory and majorization to establish a geometric reduction of the optimization over $n$-dimensional noise densities to a one-dimensional compact convex family of radially symmetric densities, whose extreme points are precisely the staircase densities.

\end{abstract}

\section{Introduction}
Differential privacy (DP)~\cite{Dwork_Calibration,Dwork-OurData} provides rigorous
privacy guarantees for statistical queries computed over sensitive datasets and has
become the standard notion of privacy in data analysis and machine learning systems
deployed in practice (e.g.,~\cite{erlingsson2014rappor,Apple_Privacy,Facebook020GuidelinesFI}).
A DP mechanism is a randomized algorithm $M$ mapping a dataset $D$ from a prescribed
collection $\bD$ to an output space $\calY$. In many applications, $M$ is implemented
by composing a deterministic query $q:\bD\to\bR^n$ with a randomized channel applied
to the query output.
Formally, a mechanism $M$ is $\eps$-DP if for all measurable sets
$A\subseteq\calY$ and all neighboring datasets $D\sim D'$ (i.e., differing in one entry), we have 
\[
\Pr(M(D)\in A) \le e^{\eps}\Pr(M(D')\in A).
\]
This ensures that the distribution of the mechanism output is insensitive to changes in any single individual’s data.

A widely used and analytically tractable class of mechanisms is \emph{additive}
mechanisms, which take the form
\[
M(D)=q(D)+X,
\]
where $X$ is an $\bR^n$-valued random vector independent of $q(D)$. 
The privacy
guarantee of such mechanisms depends on the sensitivity of the query $q$ with respect
to a given norm $\|\cdot\|$, defined as
\begin{equation}\label{eq:sensitivity}
\Delta \coloneqq \sup_{D\sim D'} \|q(D)-q(D')\|.
\end{equation}

Throughout, we restrict attention to sensitivity-calibrated additive mechanisms: the law of $X$ depends on the query $q$ only through its sensitivity $\Delta$. We refer to mechanisms of this form as \emph{query-independent additive mechanisms}.
Standard examples include the Laplace and Gaussian mechanisms, which are ubiquitous in the DP literature and applications. However, these classical mechanisms are in general suboptimal, as first observed by Hardt and Talwar~\cite{HardtGeometry}.

A central question in the theory of differential privacy is therefore the following:
\emph{among all additive noise distributions that guarantee $\eps$-DP
for every query with sensitivity at most $\Delta$, which
one minimizes utility loss?}  To formalize this question, we measure the
utility cost of an additive mechanism by $\mathbb{E}[\Phi(X)]$, where
$\Phi:\bR^n\to\bR$ is a prescribed cost function. In this work, we focus on
\emph{norm-monotone} cost functions, i.e., functions $\Phi$ satisfying $\Phi(x)\le \Phi(y)$ whenever $\|x\|\le \|y\|$.\footnote{\label{foot: norm monotone equivalence}Note that $\Phi$ is norm-monotone if and only if
$\Phi(x)=\varphi(\|x\|)$ for some nondecreasing function $\varphi$. Indeed, when $\varphi$ is an increasing function, $\Phi(x) \coloneqq  \varphi(\|x\|)$ is trivially norm increasing, so to prove the claim it is enough to show that any $\Phi$ that satisfies $\Phi(x) \leq \Phi(y)$ for $\|x\| \leq \|y\|$ has representation of the form $\Phi(x) = \varphi(\|x\|)$ for an increasing $\varphi$.  Let $e$ be such that $\|e\| = 1$ and define $\varphi(t) = \Phi(t e)$.  Since norm-monotonicity of $\Phi$ forces $\Phi(x) = \Phi(y)$ for $\|x \| = \|y\|$, $\Phi(x) = \Phi( \|x\| e)$ as $\|\|x\|e \| = \|x\|$, and by definition $\Phi(\|x\| e) = \varphi(\|x\|)$. }

This problem was solved in the scalar case ($n=1$) by Geng and Viswanath~\cite{geng2015optimal} where it was shown that an optimal
additive $\eps$-DP mechanism belongs to the family of \emph{staircase mechanisms}. Subsequently, Geng, Kairouz, Oh, and Viswanath~\cite{geng2015staircase} extended this result to $n=2$ for $\ell_1$ norm cost $\Phi(x) = \|x\|_1$. 
In these works, the staircase mechanism is defined as
\begin{equation}\label{eq:staircase_Mgamma}
M_\gamma(D) \coloneqq q(D) + X_\gamma,
\end{equation}
where $X_\gamma$ has a density of the form
\begin{equation}\label{eq:staircase-intro}
f_\gamma(x) \propto\exp\Big(-\varepsilon\Big\lfloor \tfrac{\|x\|}{\Delta}-\gamma \Big\rfloor\Big),
\qquad x\in\bR^n,
\end{equation}
for some offset parameter $\gamma\in[0,1]$. It was conjectured in
\cite{geng2015staircase} that staircase mechanisms remain optimal for all dimensions $n\ge1$ for $\ell_1$-norm cost.

More recently, Kulesza, Suresh, and Wang~\cite{Staircase_General}
considered a query-specific formulation based on a query's sensitivity space 
$$\mathcal K \coloneqq \{ q(D)-q(D') : D\sim D' \}\subseteq\bR^n.$$ When $\mathcal K$ is convex and has nonempty interior, its
Minkowski functional defines a norm $\|x\|_{\mathcal K} = \inf\{r>0 : x \in r\mathcal K\}$. They introduced the corresponding \emph{generalized} staircase
mechanism, whose noise density is proportional to
$\exp\big(-\eps\lfloor\|x\|_{\mathcal K}-\gamma\rfloor\big)$ for
$\gamma\in[0,1]$. For the query $q$, they proved that this mechanism
minimizes $\mathbb E[\|X\|_{\mathcal K}]$ among all additive
$\eps$-DP mechanisms. To recover the query-independent norm-sensitivity setting of \cite{geng2015optimal,geng2015staircase}, one may replace the exact sensitivity space by the enclosing ball
$\Delta\bB$, where $\bB\coloneqq\{u:\|u\|\leq1\}$. Since
$\|x\|_{\Delta\bB}=\|x\|/\Delta$, the generalized staircase density reduces to the staircase density in \eqref{eq:staircase-intro}. Therefore, Kulesza et al.\ establish staircase optimality in arbitrary dimension for $\Phi(x)=\|x\|$; thus resolving the conjecture of~\cite{geng2015staircase} for a specific cost function $\Phi$.

In this work, we extend this line of work in a different direction. Given an
arbitrary prescribed norm $\|\cdot\|$ and its scalar sensitivity
$\Delta$, we prove staircase optimality over the full class of
query-independent additive mechanisms for \textit{every} norm-monotone cost function $\Phi$. Thus, while the first-moment norm cost
$\Phi(x)=\|x\|$ is already covered by~\cite{Staircase_General}, our theorem
establishes optimality for the entire class of norm-monotone utility
criteria.

\begin{theorem}[Informal]\label{thm: Main}
Given a norm $\|\cdot\|$ and a norm-monotone cost function
$\Phi:\bR^n\to\bR$, there exists $\gamma^*\in[0,1]$ such that the staircase mechanism
$M_{\gamma^*}$ minimizes expected cost $\mathbb E[\Phi(X)]$ among all query-independent $\eps$-DP additive mechanisms.
\end{theorem}
Our proof relies on a novel connection between differential privacy and convex
rearrangement theory. We show that for any additive $\eps$-DP mechanism with noise
$X$, there exists another additive mechanism with noise $Y$ such that the induced
mechanism remains $\eps$-DP and satisfies $\mathbb{E}[\Phi(Y)]\le\mathbb{E}[\Phi(X)]$ (Theorem~\ref{Prop:NormRearrangementSmaller}).
Moreover, the distribution of $Y$ is radially symmetric with respect to the norm
$\|\cdot\|$ (Proposition~\ref{Prop:fastrhonorm}). This observation allows us to reduce the original infinite-dimensional optimization problem to a one-dimensional problem over a compact and convex family of radially symmetric distributions. A key technical step is the characterization of the extreme points of this family, which we show are exactly the staircase
distributions (Theorem~\ref{thm: extreme points of the space D}). This characterization leads directly to the optimality of staircase mechanisms.

\added{Our formulation differs from the query-specific sensitivity-space approach of
Kulesza et al.~\cite{Staircase_General}. Their privacy constraint is imposed
only along shifts in the exact sensitivity space $\mathcal K$, whereas our
formulation uses only the scalar norm sensitivity $\Delta$ and requires privacy
for every shift in the enclosing ball $\Delta\bB$. Thus, our formulation is
query-independent and provides a uniform guarantee for every query with
sensitivity at most $\Delta$, without requiring the full geometry of
$\mathcal K$. This can be conservative since
$\mathcal K\subseteq\Delta\bB$. For example, if
$\mathcal K=\{\pm e_1\}\subseteq\bR^2$ and
$\|\cdot\|=\|\cdot\|_2$, then noise of the form $X=(Z,0)$, with $Z$
chosen appropriately, may be
valid for the fixed query, whereas our formulation must also protect shifts
in the $e_2$ direction. Hence, the query-specific formulation may achieve
better utility for a particular query, while ours requires only the prescribed
norm and scalar sensitivity.}


\subsection{More Related Work}
Identifying optimal privacy mechanisms is a central and challenging problem in
differential privacy. In the scalar setting, a number of works have characterized
optimal mechanisms under various optimality criteria. For example, Ghosh
et al.~\cite{ghosh2012universally} showed that the geometric mechanism is universally
optimal for $\eps$-DP in a Bayesian framework, while Gupte and
Sundararajan~\cite{Gupte_universally} derived minimax-optimal noise distributions for
general cost functions. Subsequent work showed that, for a broad class of cost
functions, the optimal noise distribution in additive mechanisms exhibits a staircase-shaped
density~\cite{geng2015optimal,geng2015staircase,OptimalDP2}. More recently, Awan and Slavkovi\'c~\cite{Awan03042021} studied optimality within the class of $\mathcal K$-norm mechanisms, a family
of $\eps$-DP additive mechanisms generalizing the Laplace mechanism. These mechanisms have noise densities proportional to $\exp(-\varepsilon\|x\|_{\mathcal K})$, where $\|\cdot\|_{\mathcal K}$ is the Minkowski functional induced by a convex body $\mathcal K$. They showed that, under their comparison
criteria, the convex hull of the query's sensitivity space yields the optimal choice within this family. Thus, their work optimizes the geometry within a Laplace-type mechanism family, whereas we optimize over the full class of query-independent additive mechanisms under norm-sensitivity constraints and identify staircase mechanisms as optimal.

A complementary line of work exploits additional geometric or coordinate-wise structure of the query to improve utility. Geometry-, matrix-, and factorization-based mechanisms have been developed for linear queries, including mechanisms that introduce correlated noise across query coordinates \cite{HardtGeometry,LiMatrix,NikolovGeometry,EdmondsFactorization}.
More recently, Muthukrishnan and Kalyani \cite{MuthukrishnanKalyani} studied non-identically distributed
Laplace and Gaussian noise that exploit disparities among
coordinate-wise sensitivities. These approaches use query-specific
geometric or coordinate-wise information, whereas our framework seeks a query-independent additive mechanism calibrated only through a prescribed norm and the scalar sensitivity.

Optimal mechanism design has also been studied under weaker notions of privacy.
Geng and Viswanath~\cite{Geng_OptimalApproximate} considered $(\eps,\delta)$-DP (``approximate'' DP) for integer-valued queries and showed that discrete uniform and discrete Laplace mechanisms are asymptotically optimal for $\ell_1$ and $\ell_2$ costs in the high
privacy regime. For real-valued queries, Geng et al.~\cite{Geng_truncatedLaplace}
identified truncated Laplace distributions as asymptotically optimal in various high privacy regimes, while Geng et al.~\cite{Geng_Uniform} showed that for
$(0,\delta)$-DP the optimal noise distribution is uniform with an atom at the
origin.

In contrast to these works, which focus primarily on single-query scalar settings,
the literature on optimal mechanisms for vector-valued queries is comparatively
limited. One notable line of work studies asymptotic behavior under structural
restrictions on the noise distribution. For example,~\cite{CLT_query} showed that,
under mild regularity conditions, additive mechanisms with log-concave noise
densities behave asymptotically like the Gaussian mechanism leading to the same approximate DP guarantees. Several recent works have proposed additive mechanisms that improve upon Gaussian or Laplace noise in specific settings, including mechanisms based on normalized
Gaussian tails~\cite{OSGT_Sadeghi}, generalized Gaussian and Laplace families
\cite{vinterbo22a,GeneralizedGaussian}, and hybrid Laplace--Gaussian constructions
\cite{muthukrishnan2023grafting}. While these mechanisms can offer improved utility
in certain regimes, they are not known to be optimal. More recently,
\cite{gilani2025optimizing} formulated a numerical optimization framework for
computing optimal additive mechanisms under an alternative privacy notion, namely
R\'enyi differential privacy.

Optimal noise design has also been investigated in large composition regimes, where a
large number of queries (scalar or vector) are answered on the same dataset. In this setting, the Cactus mechanism~\cite{CactusMechanism} was shown to be optimal for approximate DP
with $\ell_2$-sensitivity, and the Schr\"odinger density~\cite{Schrodinger} was shown
to be optimal in the vanishing sensitivity regime, recovering the Gaussian
distribution as a special case under quadratic cost. Unlike these works, we focus
on the single-shot setting of a single vector-valued query under pure $\eps$-DP.

\subsection{Paper Organization}
The remainder of the paper is organized as follows. In
Section~\ref{sec: Prelims}, we formally define the problem and introduce the
necessary background on differential privacy, convex rearrangements, and stochastic
ordering. Section~\ref{sec:RearrangementToDP} develops the rearrangement-based
framework for analyzing additive $\eps$-DP mechanisms and establishes the reduction
to radially symmetric noise distributions. In Section~\ref{Sec:OptimalityStaircase}, we characterize the extreme points of the resulting feasible set and prove the optimality of staircase mechanisms. Finally, in Section~\ref{sec:samplingAlgorithm}, we propose an efficient sampling algorithm from the staircase distribution and use it to compare the staircase mechanism with Laplace mechanism. 

Most technical proofs are deferred to the appendices. Appendix~\ref{sec: ep DP is
Lipschitz Condition} establishes a basic characterization of $\eps$-DP for additive
mechanisms and shows that the corresponding noise distributions are necessarily absolutely continuous with respect to Lebesgue measure. Appendix~\ref{sec: Proof of Theorem 2} contains the main rearrangement argument underlying the reduction to radially symmetric mechanisms. Appendices~\ref{App:PropositionRadial} and \ref{sec: Lemma 1} provide proofs for the technical results connecting convex rearrangement to DP. Appendix~\ref{sec: extreme points} characterizes the extreme points of the reduced feasible set and 
finally, 
Appendix~\ref{sec: Norm Symmetric Coordinates} collects basic facts on norm-symmetric coordinates required for the proofs.

\subsection{Notation}
We work throughout on the Euclidean space $\mathbb{R}^n$, equipped with an arbitrary
norm $\|\cdot\|$. Let $\mathcal{P}(\mathbb{R}^n)$ denote the set of all Borel probability measures on $\mathbb{R}^n$, and let $m$ denote the Lebesgue measure. For a Borel set $A\subseteq\mathbb{R}^n$, we write $|A|$ for its Lebesgue measure $m(A)$.

For $x\in\mathbb{R}^n$ and $r\ge0$, define the open and closed norm balls
\[
\mathbb{B}_r(x)\coloneqq \{y\in\mathbb{R}^n : \|y-x\|<r\},
~~~\text{and}~~~~~
\overline{\mathbb{B}}_r(x)\coloneqq \{y\in\mathbb{R}^n : \|y-x\|\le r\}.
\]
We write $\mathbb{B}_r=\mathbb{B}_r(0)$, $\overline{\mathbb{B}}_r=\overline{\mathbb{B}}_r(0)$,
and $\mathbb{B}=\mathbb{B}_1$.  More generally, for $A \subseteq \mathbb{R}^n$, $\overline{A}$ denotes its closure.

For Borel sets $A,B\subseteq\mathbb{R}^n$, the Minkowski sum is
$A+B=\{a+b : a\in A,\, b\in B\}$. For a measurable function
$f:\mathbb{R}^n\to\mathbb{R}$ and $\lambda\in\mathbb{R}$, we denote the super-level
set by $\{f>\lambda\}=\{x : f(x)>\lambda\}$, and similarly define $\{f\ge\lambda\}$

and $\{f<\lambda\}$.  Throughout, ``decreasing'' is used in the weak sense: a function
$\rho$ is decreasing if $\rho(s)\ge\rho(t)$ whenever $s\le t$. We write
\emph{strictly decreasing} when the inequality is required to be strict, and
we use ``increasing'' and ``non-decreasing'' analogously.

\section{Basic Definitions and Preliminaries}\label{sec: Prelims}
In this section, we formally introduce the model, recall the definition of
$\eps$-DP for additive mechanisms, and review the notions of norm-monotone cost functions, radially symmetric rearrangements, and stochastic ordering that are used throughout the paper.
\subsection{Differential Privacy}
Let $q:\bD \to \mathbb{R}^n$ be a deterministic query, where $\bD$ is defined to be the collection of all datasets. In the absence of privacy constraints, the curator releases the query output $q(D)$. Under privacy constraints, the curator instead applies a randomized privacy mechanism
$M:\bD \to \mathcal{P}(\mathbb{R}^n)$, where $\mathcal{P}(\mathbb{R}^n)$ is the space of all Borel probability measures on $\mathbb{R}^n$.  Intuitively, $M$ satisfies differential
privacy if the distribution of its output does not change significantly when a
single individual’s record in $D$ is modified.  We will identify the probability distribution $M(D)$ with a random variable possessing the distribution $M(D)$.  More generally we have the following definition.

\begin{definition}[Differential privacy {\cite{Dwork_Calibration}}]\label{def:epsilonDP}
A randomized mechanism $M:\bD \to \mathcal Y$ satisfies
$\eps$-differential privacy ($\eps$-DP) if, for all neighboring datasets
$D \sim D'$ and all measurable sets $A$,
\[
\Pr(M(D) \in A) \le e^{\eps} \Pr(M(D') \in A).
\]
\end{definition}
We focus in this paper on \emph{additive mechanisms}, which are among the most widely
used mechanisms in practice due to their simplicity and analytical tractability.
An additive mechanism takes the form
\begin{equation}\label{eq:additiveMechanism}
M_X(D) \coloneqq q(D) + X,
\end{equation}
where $X$ is an $\mathbb{R}^n$-valued random vector independent of the dataset $D$.
The distribution of $X$ may depend on the query $q$ only through its sensitivity
with respect to a given norm $\|\cdot\|$, defined as
\[
\Delta \coloneqq \sup_{D \sim D'} \|q(D) - q(D')\|.
\]
For additive mechanisms, the $\eps$-DP constraint admits a direct characterization in terms of the noise distribution $X\sim\mu$. In particular, the query-independent additive mechanism $M_X$ satisfies $\eps$-DP if and only if
\begin{equation}\label{eq:DP01}
    \mu(A) \le e^{\eps}\mu(A + u),
\end{equation}
for all vectors $u \in \mathbb{R}^n$ with $\|u\| \le \Delta$ and all Borel sets
$A \subseteq \mathbb{R}^n$. A fundamental consequence of this condition is that $\mu$
must be absolutely continuous with respect to Lebesgue measure%
\footnote{This is a folklore result; for completeness, a proof is provided in
Theorem~\ref{thm:ExistenceDensity} in
Appendix~\ref{sec: ep DP is Lipschitz Condition}.}. In particular, via Lebesgue's differentiation theorem, the function 
\begin{align} \label{eq: explicit realization of density} f_\mu(x) \coloneqq f_X(x) \coloneqq  \limsup_{t\to 0} \frac{\mu(x + \mathbb{B}_t)}{|\mathbb{B}_t|}
\end{align}
is a density function for $\mu$.  It will follow that,
\begin{equation}\label{Eq:equivalenceAdditivedensity1}
M_X\text{ is }\varepsilon\text{-DP}
\quad \Longleftrightarrow\quad
f_X(x)\le e^{\varepsilon}f_X(y),
\end{equation}
for all $x,y\in\mathbb{R}^n$ such that $\|x-y\|\le \Delta$.  \added{Note that the inequality $f(x) \leq e^{\varepsilon} f(y)$ forces $f >0$ throughout, as $f(y) =0$ forces $f(x) = 0$ on a neighborhood of $y$ and induction implies $f = 0$, contradicting normalization.  Hence, $\log f_X$ is well defined on $\bR^n$.} \deleted{We have motivation to understand {\it{pointwise}} behavior of our density. As such} We will refer to ``the density'' associated to an $\varepsilon$-DP mechanism to be any density for $X$ that satisfies the pointwise inequality in \eqref{Eq:equivalenceAdditivedensity1}. 
Accordingly, we define
\begin{equation}\label{def:collectionkfprivatedensity}
\mathcal{M}^{\|\cdot\|}_n(\Delta,\varepsilon)
\coloneqq
\Big\{
\mu \in \mathcal{P}(\mathbb{R}^n)\;:\;
\mu \ll m, f \coloneqq f_\mu 
\ \text{satisfies}\
f(x) \le e^{\varepsilon} f(y),
\ \forall\, x,y \in \mathbb{R}^n
\ \text{with}\ \|x-y\| \le \Delta
\Big\},
\end{equation}
where $m$ denotes the Lebesgue measure. The set
$\mathcal{M}^{\|\cdot\|}_n(\Delta,\varepsilon)$ thus collects all noise distributions
that induce $\eps$-DP additive mechanisms for queries with $\|\cdot\|$-sensitivity
$\Delta$. When the norm is clear from the context, we write
$\mathcal{M}_n(\Delta,\varepsilon)$ for
$\mathcal{M}^{\|\cdot\|}_n(\Delta,\varepsilon)$.

As a canonical example, the product Laplace distribution satisfies $\eps$-DP. More
precisely, letting $\mathcal L(0,b)$ denote the zero-mean Laplace distribution with
scale parameter $b$, one can verify that
$\mathcal L^{\otimes n}(0,\Delta/\varepsilon)\in
\mathcal{M}_n(\Delta,\varepsilon)$ for $\|\cdot\|=\|\cdot\|_1$.
Another important example is the \emph{staircase mechanism}~\cite{geng2015optimal},
which we describe next.  Throughout, $\mathbbm{1}_A$ denotes the indicator function of a set $A$, that is, for a set $A$, $$\mathbbm{1}_A(x) = \begin{cases}
    1 &\mbox{ when } x \in A
    \\
    0 &\mbox{ when } x \notin A.
\end{cases}$$

\begin{definition}[Staircase mechanism {\cite{geng2015optimal,geng2015staircase}}]
\label{def:Staircase_mech}
Let $\gamma\in[0,1]$ and let $q$ be a query with $\|\cdot\|$-sensitivity $\Delta$.
The \emph{staircase mechanism} is defined as
\[
M_\gamma(D) \coloneqq q(D) + X_\gamma,
\]
where $X_\gamma$ is a random vector with density
\begin{align*}
f_\gamma(x)
&\coloneqq a(\gamma)
\sum_{k=0}^\infty e^{-k\varepsilon}
\mathbbm{1}_{[(k-1+\gamma)\Delta,(k+\gamma)\Delta)}(\|x\|)\\
&=
\begin{cases}
a(\gamma)e^{-k\varepsilon}, & \|x\|\in[k\Delta,(k+\gamma)\Delta),\\[2pt]
a(\gamma)e^{-(k+1)\varepsilon}, & \|x\|\in[(k+\gamma)\Delta,(k+1)\Delta),
\end{cases}
\end{align*}
with $a(\gamma)$ denoting the normalizing constant.
\end{definition} 
Note{\footnote{ Indeed, for $x \in [(k-1) \Delta,  k\Delta)$ so that
\(
f_0(x) = a(0) e^{-k \varepsilon}
\)
Meanwhile considering $\gamma =1$, the same interval containing $x$ can be expressed as  $[(k-2 +1)\Delta, (k-1 + 1)\Delta)$ so that $f_1(x) = a(1) e^{- (k-1)\varepsilon}$. In particular
$ f_1(x) = \tfrac{a(1)e^{\varepsilon}}{a(0)} f_0(x)$.  Integrating the equality gives $\tfrac{a(1)e^{\varepsilon}}{a(0)} =1$ and $f_0 = f_1$.}} $f_0 = f_1$
and observe that for any $\gamma$, the density $f_\gamma$ takes exactly \emph{two} distinct values on each
interval $\|x\|\in[k\Delta,(k+1)\Delta]$ for $k\in\mathbb{N}$—one proportional to
$e^{-k\varepsilon}$ and the other to $e^{-(k+1)\varepsilon}$.  Moreover, for any
$x,y\in\mathbb{R}^n$ satisfying $\|x-y\|\le\Delta$, with $\| x\| \leq \|y \| \in [(k + \gamma) \Delta, (k+1 +\gamma) \Delta)$ for concreteness, we have from the triangle inequality,
$\|y\|-\|x\|\le\Delta$, and hence  $\|x\| \geq (k-1+\gamma) \Delta$ so that
\[
    0 \leq \log f_\gamma(x) - \log f_\gamma(y) \leq (- k \varepsilon) - (-(k+1)\varepsilon) = \varepsilon. 
\]
Consequently,
$f_\gamma(x)\le e^\varepsilon f_\gamma(y)$, showing that $M_\gamma$ satisfies
$\varepsilon$-DP for all $\gamma\in[0,1]$. 

It was shown in~\cite{geng2015staircase} that, in the two-dimensional case ($n=2$), a particular choice of $\gamma$ minimizes the $\ell_1$ cost among all additive $\eps$-DP mechanisms.
\begin{theorem}[{Geng et~al. \cite[Theorem 2]{geng2015staircase}}]
\label{thm:Geng_2d}
For queries with $\|\cdot\|_1$-sensitivity $\Delta$, the $\varepsilon$-DP additive mechanism with
minimal $\ell_1$ cost is the staircase mechanism $M_{\gamma^\ast}$, where
\[
\gamma^\ast \coloneqq \argmin_{\gamma\in[0,1]}
\int_{\mathbb{R}^2}\|x\|_1 f_\gamma(x)\,\mathrm{d}x.
\]
\end{theorem}
In particular, Theorem~\ref{thm:Geng_2d} implies that
\[
\inf_{\mu\in \mathcal{M}_2(\Delta, \varepsilon)}
\int_{\mathbb{R}^2} \|x\|_1 \mu(\mathrm{d}x)
=
\inf_{\gamma \in [0,1]}
\int_{\mathbb{R}^2} \|x\|_1 f_\gamma(x)\,\mathrm{d}x,
\]
thereby reducing the original infinite-dimensional optimization over
$\varepsilon$-DP densities to a one-dimensional optimization over the parameter
$\gamma$. This result extends \cite[Theorem~1]{geng2015optimal}, which establishes
the optimality of the one-dimensional staircase mechanism ($n=1$) for a broader
class of cost functions.

Building on these results, Geng et~al.~\cite{geng2015staircase} conjectured that the
staircase mechanism remains optimal for all dimensions $n\ge1$ under $\ell_1$ cost. This conjecture has been recently proved by Kulesza et al.~\cite{Staircase_General} for a specific choice of cost function. 
In this work, we confirm this conjecture for a broader class of cost functions, including all norm-monotone costs. More specifically, we aim to
characterize the noise distribution
$\mu\in\mathcal{M}_n(\Delta,\varepsilon)$ minimizing the expected cost
$\mathbb{E}_\mu[\Phi(X)]$ for a given cost function
$\Phi:\mathbb{R}^n\to\mathbb{R}$. 
Our objective is therefore to characterize
\begin{equation}\label{eq:CostOptimization}
\inf_{\mu\in\mathcal{M}_n(\Delta,\varepsilon)}
\int_{\mathbb{R}^n}\Phi(x)\,\mu(\mathrm{d}x).
\end{equation}
We fully characterize the minimizer of~\eqref{eq:CostOptimization} for all $n$ when
$\Phi$ is norm-monotone.
\begin{assumption}\label{assm:norm_monotone}
The cost function $\Phi:\mathbb{R}^n\to[0,\infty]$ is of the form
\[
\Phi(x)=\varphi(\|x\|),
\]
for some nondecreasing function $\varphi:[0,\infty)\to[0,\infty]$. 
\end{assumption}

Note from Footnote \ref{foot: norm monotone equivalence}, $\Phi(x) \leq \Phi(y)$ for $\|x\|  \leq \|y\|$. Typical examples of norm-monotone cost functions include the mean squared error $\Phi(x) = \|x\|^2$ and more generally $\Phi(x)=\|x\|^p$ for
$p>0$, indicator-type losses of the form
$\Phi(x)=\mathbbm{1}_{\{\|x\|\ge \lambda\}}$ for $\lambda>0$ which capture
threshold-based performance criteria, and truncated growth $\Phi(x)=\min\{\|x\|,T\}$ for some $T\geq 0$.

\subsection{Radially Symmetric Decreasing Rearrangements}
For a measurable set $A \subseteq \mathbb{R}^n$, let $|A|$ denote its Lebesgue
measure. The radially symmetric decreasing rearrangement of a set replaces $A$ by a
centered $\|\cdot\|$-ball of equal volume.
\begin{definition}\label{Def:ConvexRearrangement}
For a measurable set $A \subseteq \mathbb{R}^n$, define its rearrangement as
\[
A^\ast \coloneqq
\Big\{
x \in \mathbb{R}^n :
\|x\| <
\big(|A|/|\mathbb{B}|\big)^{1/n}
\Big\},
\]
where $\mathbb{B}$ denotes the unit ball induced by the norm $\|\cdot\|$.
\end{definition}
Thus, $A^\ast$ is the open centered $\|\cdot\|$-ball having the same Lebesgue measure
as $A$. Rearrangements extend naturally to nonnegative functions by rearranging their level
sets. Indeed, for any measurable function
$f:\mathbb{R}^n \to [0,\infty)$, we may write
$f(x)=\int_0^\infty \mathbbm{1}_{\{f(x)>\lambda\}}\,\mathrm{d}\lambda$ and rearrange the
super-level sets.

\begin{definition}\label{Def:ConvexRearrangement2}
Let $f:\mathbb{R}^n \to [0,\infty)$ be measurable. Its radially symmetric decreasing
rearrangement $f^\ast:\mathbb{R}^n \to [0,\infty)$ is defined by
\[
f^\ast(x)
\coloneqq
\int_0^\infty
\mathbbm{1}_{\{f>\lambda\}^\ast}(x)\,\mathrm{d}\lambda.
\]
\end{definition}

By construction, the level sets of $f$ and $f^\ast$ satisfy an exact set-theoretic identity
$\{f^\ast>\lambda\}=\{f>\lambda\}^\ast$ for all $\lambda \in \mathbb{R}$, which follows from the fact that both sets are open origin-symmetric balls of the same measure, see Proposition~\ref{Prop:PropertiesRearrangements} Item~\ref{item: set-theoretic equality} below for a rigorous proof. In particular, $f$ and

$f^\ast$ are equimeasurable, i.e.,
$|\{f^\ast>\lambda\}|=|\{f>\lambda\}|$ for every $\lambda>0$. Consequently,
\begin{align*}
\int_{\mathbb{R}^n} f(x)\,\mathrm{d}x
&= \int_0^\infty |\{f>\lambda\}|\,\mathrm{d}\lambda
\\ & = \int_0^\infty |\{f^\ast>\lambda\}|\,\mathrm{d}\lambda \\
&= \int_{\mathbb{R}^n} f^\ast(x)\,\mathrm{d}x,
\end{align*}
confirming that if $f$ is a probability density on $\mathbb{R}^n$, then so is
$f^\ast$. This motivates the following definition.
\begin{definition}\label{Def:ConvexRearrangement3}
Let $X$ be an $\mathbb{R}^n$-valued random vector with density $f$. The
rearrangement of $X$, denoted by $X^\ast$, is the random vector with density
$f^\ast$.
\end{definition}

We summarize several properties of radially symmetric decreasing rearrangements that
are used throughout the paper.

\begin{proposition}\label{Prop:PropertiesRearrangements}
For measurable sets $A,B \subseteq \mathbb{R}^n$ and measurable
$f:\mathbb{R}^n \to [0,\infty)$, the following hold:
\begin{enumerate}[label=(\roman*)]
\item $|A^\ast| = |A|$.
\item $\{f^\ast>\lambda\} = \{f>\lambda\}^\ast$ for all $\lambda \in \mathbb{R}$. In particular $f^*$ is lower semicontinuous.\label{item: set-theoretic equality}
\item $|A^\ast \cap (B^\ast)^c| \le |A \cap B^c|$.
\item $|A^\ast\cap B^\ast|\;\ge\;|A\cap B|.$ 
\item $|A^\ast + B^\ast| \le |A + B|$.
\item If $\Psi:\mathbb{R}_+ \to \mathbb{R}$ is strictly increasing and continuous,
then $\Psi\circ f^\ast = (\Psi\circ f)^\ast$.
\end{enumerate}
\end{proposition}

The proof of Proposition~\ref{Prop:PropertiesRearrangements} is given in
Appendix~\ref{sec: Proof of Theorem 2}.

\subsection{Stochastic Domination}

We recall a standard stochastic ordering that will be used to compare the utility
costs induced by different noise distributions; see
\cite{marshall1979inequalities,shaked2007stochastic} for background and further
discussion.

\begin{definition}
Let $U$ and $V$ be non-negative\footnote{The non-negativity in the variables is non-essential and stochastic domination can be stated analogously for general random variables, see \cite{shaked2007stochastic}.} random variables. We say that $V$
\emph{stochastically dominates} $U$, and write $U \prec V$, if
\[
\Pr(U>\lambda) \le \Pr(V>\lambda),
\]
for all $\lambda \ge 0$.
\end{definition}

Since $\mathbb{E}[U]=\int_0^\infty \Pr(U>t)\,dt$ for any non-negative random variable
$U$, it follows immediately that $U \prec V$ implies
$\mathbb{E}[U] \le \mathbb{E}[V]$. The converse does not hold in general. However, a
well-known characterization of stochastic domination states that
\begin{equation}\label{eq:equivalenceDomination}
U \prec V
\quad \Longleftrightarrow \quad
\mathbb{E}[\varphi(U)] \le \mathbb{E}[\varphi(V)],
\end{equation}
for all increasing functions
$\varphi:[0,\infty)\to\mathbb{R}$; see~\cite{shaked2007stochastic}.

\section{From Rearrangement to Differential Privacy}\label{sec:RearrangementToDP}
In this section, we establish a precise connection between differential
privacy and radially symmetric decreasing rearrangements. This connection provides
the main technical framework used in the subsequent section to prove our main result. We first characterize $\eps$-DP additive mechanisms in terms of structural properties of their noise densities. We then show how this characterization can be expressed as a geometric inclusion property of density level sets, which allows rearrangement theory to be applied.

\begin{definition}
    For a set $A \subseteq \mathbb{R}^n$ and a scalar $h \ge 0$, define the $h$-enlargement of $A$
with respect to a metric $d$ as
\[
 A_{d,h} \coloneqq 
\{x \in \mathbb{R}^n : d(x,y) < h \text{ for some } y \in A\}.
\]
\end{definition}
When the choice of metric is clear we will drop the subscript $d$, and write simply $A_h$ instead of $A_{d,h}$.

\begin{theorem}
\label{Thm:DPequivalentEnlarged}
For an $\mathbb{R}^n$ valued random vector, the induced additive mechanism $M_X$, achieves $\varepsilon$-DP when added to any query with $\|\cdot\|$-sensitivity $\Delta$ if and only if its noise density $f_X$ satisfies
\begin{equation}
\label{eq:dpenlargement}
\{f_X > \lambda\}_h
\subseteq
\{f_X > \lambda e^{-h}\},
\end{equation}
for all $\lambda>0$ and $h>0$, where $\{f_X > \lambda\}_h$ denotes the $h$-enlargement of $\{f_X > \lambda\}$ with
respect to the metric
\[
d_{\varepsilon,\Delta}(x,y)
\coloneqq
\varepsilon \Big\lceil \tfrac{\|x-y\|}{\Delta} \Big\rceil .
\]
\end{theorem}

The proof of this theorem is based on two observations.
First, using~\eqref{Eq:equivalenceAdditivedensity1}, one shows that $M_X$ is
$\varepsilon$-DP if and only if $\log f_X$ is $1$-Lipschitz with respect to the metric
$d_{\varepsilon,\Delta}$, i.e.,
\[
|\log f_X(x) - \log f_X(y)|
\le d_{\varepsilon,\Delta}(x,y),
\]
for all $x,y\in\mathbb{R}^n$.
Second, for a general function $f$, the property that $\log f$ is $1$-Lipschitz with
respect to a metric $d$ is equivalent to the level-set inclusion
$\{f>\lambda\}_h \subseteq \{f>\lambda e^{-h}\}$ for all $\lambda,h>0$. Applying this
equivalence with $d=d_{\varepsilon,\Delta}$ yields~\eqref{eq:dpenlargement}. A complete
proof is given in Appendix~\ref{sec: ep DP is Lipschitz Condition}.

Theorem~\ref{Thm:DPequivalentEnlarged} provides a geometric interpretation of privacy:
the super-level sets of a DP noise density expand in a controlled manner under the
metric $d_{\varepsilon,\Delta}$. This characterization allows rearrangement arguments
to be applied directly, leading to the following key result.

\begin{theorem}
\label{Prop:NormRearrangementSmaller}
If $M_X$ is $\varepsilon$-DP, then the rearranged mechanism $M_{X^\ast}$, where
$X^\ast$ denotes the rearrangement of $X$, is also
$\varepsilon$-DP. Moreover,
\[
\|X^\ast\| \prec \|X\|.
\]
\end{theorem}
To prove this theorem, we first reformulate the $\varepsilon$-DP constraint as a geometric inclusion property of density level sets using Theorem~\ref{Thm:DPequivalentEnlarged}. Exploiting monotonicity and set-inclusion properties of radially symmetric decreasing rearrangements, we show that these inclusions are preserved under rearrangement, implying that the rearranged mechanism remains $\varepsilon$-DP. A majorization argument on the radial profiles then yields the above stochastic domination. A complete proof is given in
Appendix~\ref{sec: Proof of Theorem 2}. 

Combined with the stochastic domination characterization~\eqref{eq:equivalenceDomination}, this theorem implies that rearranging can only reduce the expected cost while preserving privacy. In particular, it demonstrates that
$\mathbb{E}[\Phi(X^\ast)] \le \mathbb{E}[\Phi(X)]$ for any norm-monotone cost function $\Phi$.  
To exploit this structure further, we characterize the form of the rearranged density.
\begin{proposition}
\label{Prop:fastrhonorm}
Let $f$ be a probability density on $\mathbb{R}^n$, and let $f^\ast$ denote its
rearrangement. Then there exists a decreasing \added{lower semi-continuous} function
$\rho:[0,\infty)\to[0,\infty)$ such that
\begin{equation}
\label{eq:fastrhonorm}
f^\ast(x) = \rho(\|x\|),
\end{equation}
for all $x\in\mathbb{R}^n$. Moreover, $f^\ast$ satisfies the DP constraint
\eqref{eq:dpenlargement} if and only if $\log \rho$ is $1$-Lipschitz with respect to the
one-dimensional metric
\[
d_{\varepsilon,\Delta}^{(1)}(r,s)
\coloneqq
\varepsilon \Big\lceil \tfrac{|r-s|}{\Delta} \Big\rceil .
\]

\end{proposition}
Proof of this proposition is given in Appendix~\ref{App:PropositionRadial}.  
Proposition~\ref{Prop:fastrhonorm} shows that any rearranged DP density is fully
characterized by a one-dimensional radial profile $\rho$. Consequently, the
optimization problem~\eqref{eq:CostOptimization} over $n$-dimensional densities in
$\mathcal{M}_n(\Delta,\varepsilon)$ reduces to a one-dimensional problem over $\rho$.
In the next section, we study this reduced formulation to characterize the optimal
noise distribution.

\section{Optimality of the Staircase Mechanism}
\label{Sec:OptimalityStaircase}

In this section, we build on the connection between differential privacy and
rearrangement developed in the previous section to establish the
optimality of the staircase mechanism.

Recall from Theorem~\ref{Prop:NormRearrangementSmaller} that, for any additive
$\varepsilon$-DP mechanism, rearranging the noise yields another $\varepsilon$-DP
mechanism with no larger expected cost. In particular, any admissible noise
distribution may be symmetrized without loss of privacy or utility. Moreover,
Proposition~\ref{Prop:fastrhonorm} shows that the resulting noise density is radially
symmetric. Together, these results imply that any optimal noise distribution must
belong to a structured subset of $\mathcal{M}_n(\Delta,\varepsilon)$. We next show
that optimal noise distributions enjoy an additional extremal property, allowing the
optimization problem to be further simplified. To formally describe this extremal property, we define the following. 
\begin{definition}
\label{def: The space D}
Let $\mathcal{D} \subseteq \mathcal{M}_n(\Delta, \varepsilon)$ denote the collection
of probability measures on $\mathbb{R}^n$ that possess  a density with respect to the Lebesgue measure, $f(x)=\rho(\|x\|)$, for some $\rho:[0,\infty) \to \mathbb{R}_+$ decreasing and lower semi-continuous and satisfying
the maximal decay condition
\begin{equation}
\label{eq: maximal decrease}
\rho(t+\Delta)=e^{-\varepsilon}\rho(t),
\end{equation}
for all $t\geq 0$.
\end{definition}
In the following lemma, we show the existence of an optimal noise distribution belonging to $\mathcal{D}$.  

\begin{lemma}
\label{lem: X majorizes a Y in D}
For any random vector $X \sim \mu \in \mathcal{M}_n(\Delta, \varepsilon)$, there exists
a random vector $Y$ with distribution in $\mathcal{D}$ such that
\[
\|Y\| \prec \|X\|.
\]
\end{lemma}
The proof of this lemma is given in Appendix~\ref{sec: Lemma 1}. Compared to Proposition~\ref{Prop:fastrhonorm}, this lemma allows us to restrict the search for optimal noise distributions to the much smaller set $\mathcal{D}$. Crucially, $\mathcal{D}$ is convex and weakly compact, and its extreme points admit a simple and familiar characterization.

\begin{theorem}
\label{thm: extreme points of the space D}
The set $\mathcal{D}$ is convex and compact in the weak topology. Moreover, its
extreme points are exactly the staircase distributions.
\end{theorem}
The proof of Theorem~\ref{thm: extreme points of the space D} is provided in
Appendix~\ref{sec: extreme points}. As a consequence of this theorem and the
Krein--Milman theorem\footnote{See Appendix~\ref{sec: extreme points} for a simple version of the Krein-Milman theorem.}, any distribution in $\mathcal{D}$ can be approximated in the
weak topology by finite convex combinations of staircase distributions. This directly leads to our main optimality result, delineated next. 
\begin{theorem}\label{thm:conjecture_resolution}
Let $\varepsilon>0$, $\Delta>0$, and let $\Phi$ satisfy
Assumption~\ref{assm:norm_monotone}.
Then staircase distributions are optimal noise distributions in the sense that
\begin{equation}\label{eq:inf_equalities}
\inf_{\mu\in\mathcal{M}_n(\Delta,\varepsilon)}
\int_{\mathbb{R}^n}\Phi(x)\,\mu(\mathrm{d}x)
=
\inf_{\gamma\in[0,1]}
\int_{\mathbb{R}^n}\Phi(x)\,f_\gamma(x)\,\mathrm{d}x,
\end{equation}
where $f_\gamma$ is a staircase density as in Definition \ref{def:Staircase_mech}.
Moreover, the infimum on the right-hand side is attained: there exists
$\gamma^\ast\in[0,1]$ such that
\begin{equation}\label{eq:Gamma^*}
\inf_{\gamma\in[0,1]} \int_{\mathbb{R}^n}\Phi(x)\,f_\gamma(x)\,\mathrm{d}x = \int_{\mathbb{R}^n}\Phi(x)\,f_{\gamma^\ast}(x)\,\mathrm{d}x,
\end{equation}
and the staircase mechanism $M_{\gamma^\ast}$ is optimal among all $\varepsilon$-DP \added{query-independent} additive mechanisms.
\end{theorem}

{The proof uses some standard definitions from functional analysis, which we now recall.  The typical framework is that of a locally convex Hausdorff vector space, $E$ which is to say that the topology of $E$ is Hausdorff can be generated by a base of convex sets where the vector space operations of addition and scalar multiplication are continuous functions. Our investigations take place on the space of all Borel probability measures on $\mathbb{R}^n$, $\mathcal{P}(\mathbb{R}^n)$, which when viewed as a subset of $C_b(\mathbb{R}^n)^*$, the dual space to $C_b(\mathbb{R}^n)$, the continuous bounded functions on $\mathbb{R}^n$ with the weak-$*$ topology, gives exactly the so called topology of ``weak-convergence'' in the terminology of probability.  For general set $A \subseteq E$, its convex hull is defined by considering $\mathcal{K_A} \coloneqq \{ K \subseteq E: K \mbox{ is convex}, A \subseteq K \}$ and defining
\[
    co(A) = \bigcap_{K \in \mathcal{K}_A} K.
\]
The closed convex hull of $A$, denoted $\overline{co}(A)$ is the topological closure of $co(A)$. Finally a point $x$ of a convex set $K$, is called an extreme point if $x = (1-t)x_1 + t x_2$ for two points $x_i \in K$ and $t \in (0,1)$ implies that $x_1 = x_2 = x$.  The set of all extreme points of $K$ will be denoted by  $\mathcal{E}(K)$.}

\begin{proof}
Let $X\sim\mu\in\mathcal{M}_n(\Delta,\varepsilon)$. By
Lemma~\ref{lem: X majorizes a Y in D}, there exists
$Y\sim\nu\in\mathcal D$ such that $\|Y\|\prec \|X\|$.
Since $\Phi(x)=\varphi(\|x\|)$ with $\varphi$ nondecreasing
(Assumption~\ref{assm:norm_monotone}),
the stochastic domination characterization~\eqref{eq:equivalenceDomination} yields
\[
\mathbb{E}[\Phi(Y)]
=\mathbb{E}[\varphi(\|Y\|)]
\le \mathbb{E}[\varphi(\|X\|)]
=\mathbb{E}[\Phi(X)].
\]
Taking infima over $\mu\in\mathcal{M}_n(\Delta,\varepsilon)$ gives
\[
\inf_{\nu\in\mathcal D}\int \Phi\, \mathrm{d}\nu
\le
\inf_{\mu\in\mathcal{M}_n(\Delta,\varepsilon)}\int \Phi\,\mathrm{d}\mu.
\]
Since $\mathcal D\subseteq\mathcal{M}_n(\Delta,\varepsilon)$, the reverse inequality
holds trivially, and hence
\begin{equation}\label{eq:reduce_to_D_clean}
\inf_{\mu\in\mathcal{M}_n(\Delta,\varepsilon)}\int \Phi\,\mathrm{d}\mu
=
\inf_{\nu\in\mathcal D}\int \Phi\,\mathrm{d}\nu.
\end{equation}

{Thus, if $\int \Phi \,\mathrm{d}\nu = \infty$ for all $\nu \in \mathcal{D}$, then $F: \mathcal{D} \to [0,\infty]$ defined as $F(\mu) \coloneqq \int \Phi \mathrm{d}\mu$ is identically infinity and there is nothing to prove. Hence,  we can assume that there exists $\mu' \in \mathcal{D}$ such that $F(\mu') < \infty$ and hence by Corollary~\ref{cor: continuity of F},  $F$ is continuous.  Consequently,  $F(\mathcal{D})$ is a compact set as the continuous image of a compact set and there exists $\mu \in \mathcal{D}$ such that $F(\mu)$ is minimal.}
We next reduce the optimization over $\mathcal D$ to its extreme points.  
{ By Krein-Milman, $\mathcal{D} = \overline{co}(\mathcal{E}(\mathcal{D}))$, and therefore there exists a sequence $\mu_k \in co(\mathcal{E}(\mathcal{D}))$ converging to $\mu$.  For any element $\tilde{\mu}$ belonging to the convex hull, there exists $\nu_i \in \mathcal{E}(\mathcal{D})$ and $\lambda_i \geq 0$ such that $\sum_{i=1}^m \lambda_i =1$ and $\sum_{i=1}^m \lambda_i \nu_i = \tilde{\mu}$. We can then write 
\begin{align*}
    F(\tilde{\mu})
        = 
            F \left( \sum_{i=1}^m \lambda_i \nu_i \right)
        =
            \sum_{i=1}^m \lambda_i F( \nu_i)
        \geq
            \inf_{\nu \in \mathcal{E}(\mathcal{D})} F(\nu).
\end{align*}
Thus, applying this bound to $\mu_k$, alongside the continuity of $F$, implies 
\begin{align*}
    F(\mu) = \lim_{k \to \infty} F(\mu_k) \geq \lim_{k \to \infty} \left(\inf_{\nu \in \mathcal{E}(\mathcal{D})} F(\nu) \right) =\inf_{\nu \in \mathcal{E}(\mathcal{D})} F(\nu) = \inf_{\gamma\in[0,1]}\int_{\mathbb R^n}\Phi(x)\,f_\gamma(x)\,\mathrm{d}x.
\end{align*}
}

Finally, the continuity of $F$ \added{and the weak continuity of $\gamma\mapsto\mu_\gamma$, including at $f_0=f_1$ (which follows from the a.e.\ continuity of $f_\gamma$ in $\gamma$ and dominated convergence),} implies that $\gamma\mapsto\int \Phi(x)\,f_\gamma(x)\,\mathrm{d}x$ is continuous
on the compact set $[0,1]$.  Thus, it attains its infimum at some
$\gamma^\ast\in[0,1]$, proving~\eqref{eq:Gamma^*} and the optimality of
the staircase mechanism $M_{\gamma^\ast}$.
\end{proof}

\section{Sampling and Numerical Illustrations}\label{sec:samplingAlgorithm}

In this section, we address the implementation and numerical evaluation of the staircase mechanism. We first give a sampling procedure for the staircase density $f_\gamma$ under an arbitrary norm $\|\cdot\|$ for which a uniform
sampler for its unit ball is available. It provides a practical method for generating staircase noise in private releases and downstream procedures. We then specialize the construction to the $\ell_1$- and $\ell_2$-norms.

For the radial costs considered in our numerical comparisons, sampling is not needed to evaluate the objective. Since both $f_\gamma$ and $\Phi(x)=\varphi(\|x\|)$ depend only on the radius, $J(\gamma)=\mathbb E_{f_\gamma}[\Phi(X)]$ reduces to a deterministic one-dimensional radial series. We use this representation to optimize over the scalar parameter $\gamma\in[0,1]$. The resulting figures are intended only as numerical illustrations of the magnitude of the utility gains guaranteed by our theoretical results.

\begin{figure}[t]
\centering
    \includegraphics[scale=0.13]{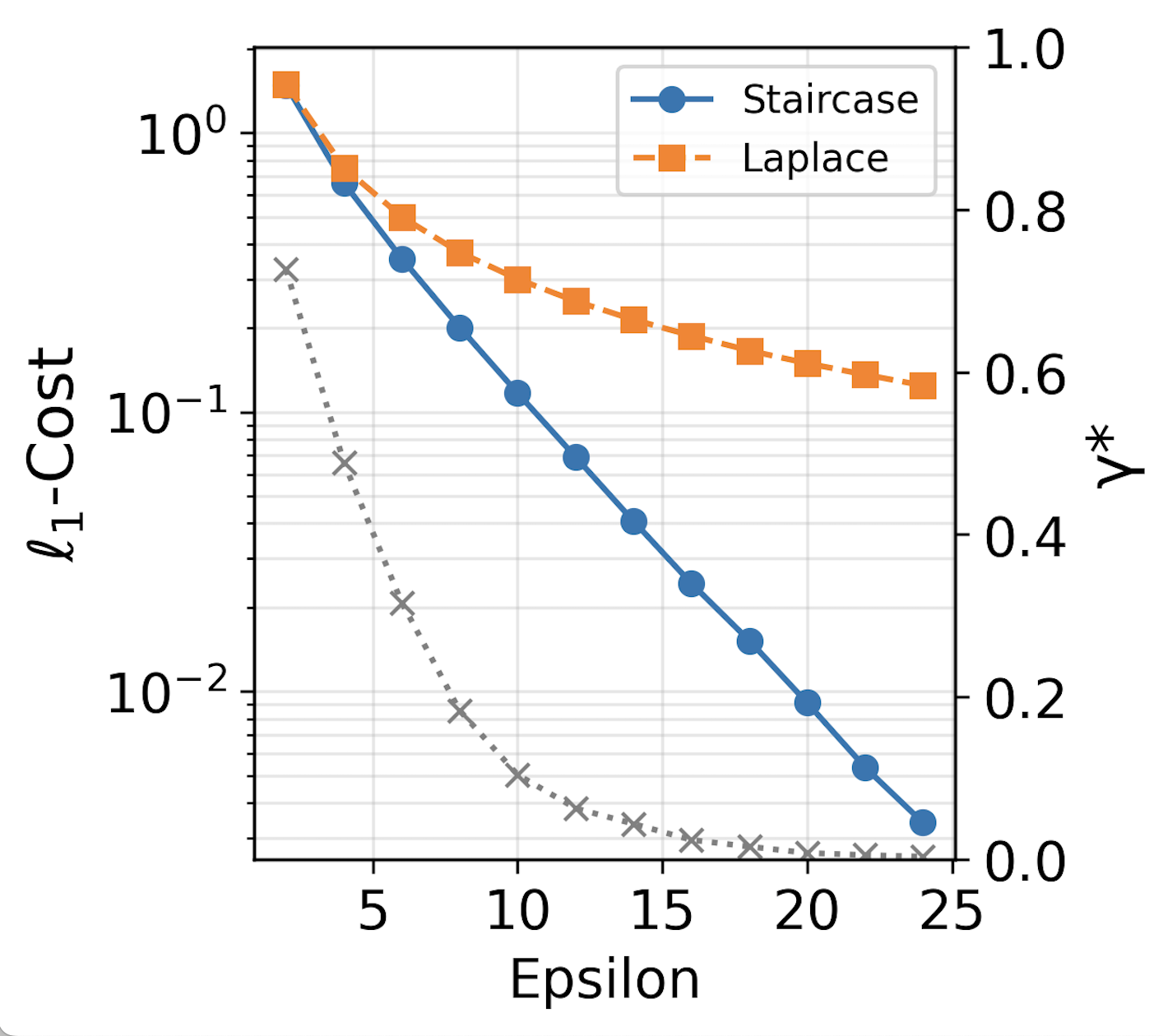}\qquad~~~~ \includegraphics[scale=0.13]{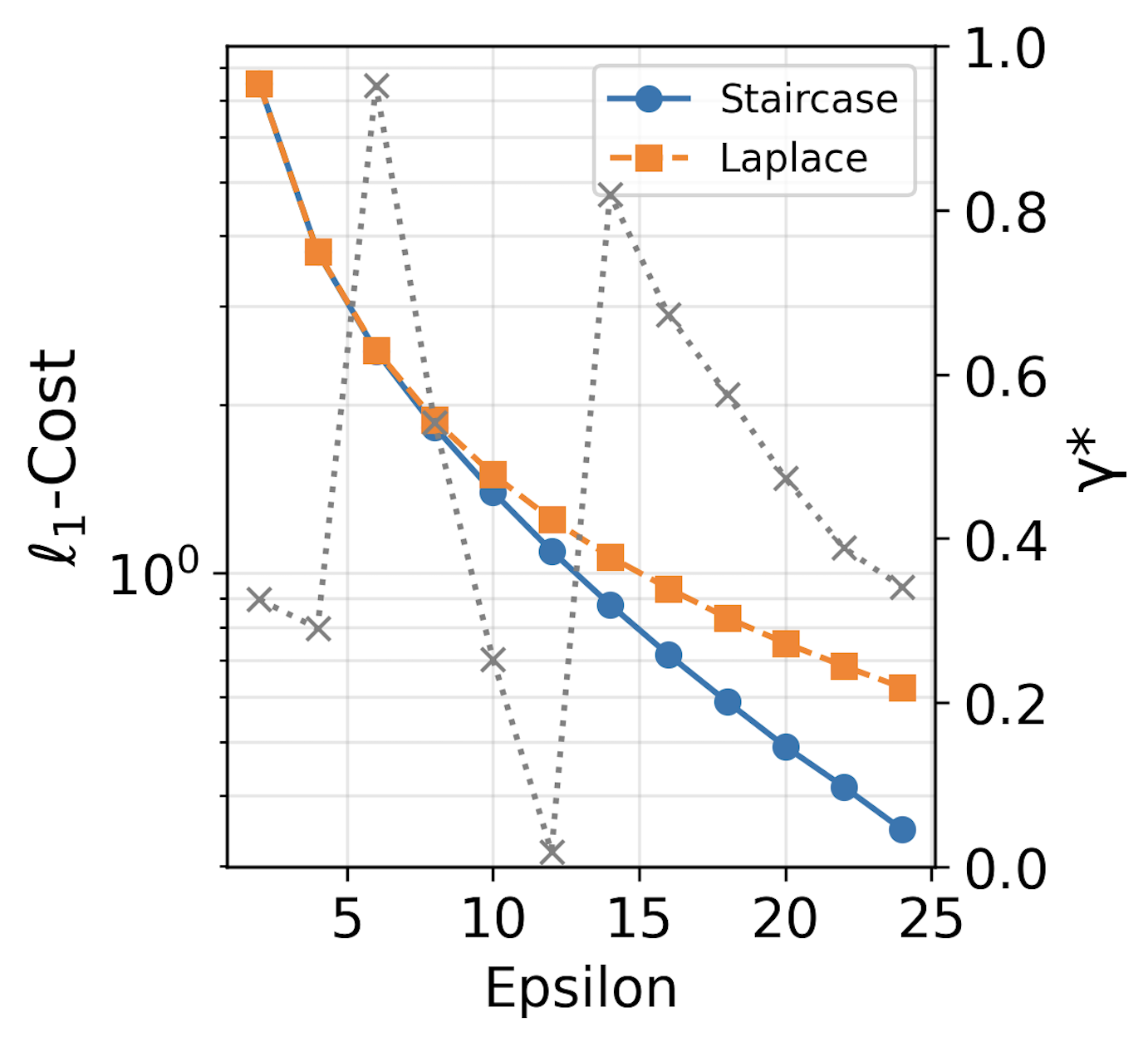}
    \caption{The utility-privacy tradeoff for staircase mechanism and Laplace mechanism for $n = 3$ (left) and $n= 15$ (right) with $\|\cdot\|_1$-sensitivity one, where the utility is measured via the cost function $\Phi(x) = \|x\|_1$. The gray dotted curve shows the optimal $\gamma^\ast$, the solution of the optimization problem \eqref{eq:Gamma^*}.}
    \label{fig:Delta1}
\end{figure}
\subsection{Analytical Properties of the Staircase Density}
Recall that the staircase density $f_\gamma$ is radially symmetric and piecewise
constant in the radius $\|x\|$, with step width $\Delta$ and exponentially decaying
heights. For each integer $k \ge 0$, the density takes two constant values on the
interval $\|x\| \in [k\Delta,(k+1)\Delta]$, with a breakpoint at $(k+\gamma)\Delta$.
Define the two sub-intervals
$$
B_{k,1} \coloneqq [k\Delta,(k+\gamma)\Delta),$$
and $$B_{k,2} \coloneqq [(k+\gamma)\Delta,(k+1)\Delta).$$
Note that the density $f_\gamma$ is constant on each such band. 
Let $V(r)\coloneqq \mathrm{Vol}(\overline{\mathbb{B}}_r)$ denote the volume of the
closed ball of radius $r$ under the norm $\|\cdot\|$, and define
\[w_{k,i} \coloneqq \Pr(\|X\| \in B_{k,i}), \]
for $i\in\{1,2\}$. By construction, we have 
\begin{align*}
w_{k,1}
&= a(\gamma)\,e^{-k\varepsilon}
\Big(V\big((k+\gamma)\Delta\big)-V\big(k\Delta\big)\Big),\\
w_{k,2}
&= a(\gamma)\,e^{-(k+1)\varepsilon}
\Big(V\big((k+1)\Delta\big)-V\big((k+\gamma)\Delta\big)\Big),
\end{align*}
where $a(\gamma)$ is the normalizing constant in the staircase density.
By the homogeneity of the norm,
$V(r)=V(1)r^n$. Therefore, up to a common normalizing factor,
\begin{align*}
w_{k,1}
&\propto
e^{-k\eps}\big((k+\gamma)^n-k^n\big),\\
w_{k,2}
&\propto
e^{-(k+1)\eps}
\big((k+1)^n-(k+\gamma)^n\big).
\end{align*}
For $m>0$, define
$$
S_m(\gamma)
\coloneqq
\sum_{k=0}^{\infty}
\left[
e^{-k\eps}\big((k+\gamma)^m-k^m\big)
+
e^{-(k+1)\eps}
\big((k+1)^m-(k+\gamma)^m\big)
\right].
$$
The normalizing constant is therefore
$$
a(\gamma)
=
\frac{1}{V(1)\Delta^n S_n(\gamma)}.
$$
Consequently, for every radial cost
$\Phi(x)=\varphi(\|x\|)$,
\begin{align}
J(\gamma)
&=
\frac{n}{S_n(\gamma)}
\sum_{k=0}^{\infty}
\bigg[
e^{-k\eps}
\int_k^{k+\gamma}
\varphi(\Delta t)t^{n-1}\,dt
+
e^{-(k+1)\eps}
\int_{k+\gamma}^{k+1}
\varphi(\Delta t)t^{n-1}\,dt
\bigg].
\label{eq:radial-cost}
\end{align}
Thus, evaluating $J(\gamma)$ requires only one-dimensional integrals and involves no Monte Carlo estimation. In particular, for $\varphi(r)=r^p$ with $p>0$,
\begin{equation}
\mathbb E_{f_\gamma}\!\left[\|X\|^p\right]
=
\frac{n}{n+p}\Delta^p
\frac{S_{n+p}(\gamma)}{S_n(\gamma)},
\label{eq:staircase-moment}
\end{equation}
which we will use later. 

\subsection{Two-Stage Sampling Procedure}
We wish to generate a sample $X$ from density $f_\gamma$ for a given $\gamma\in [0, 1]$. 
A direct sampling from $f_\gamma$ proceeds in two stages:

\begin{itemize}
\item \emph{Band selection:}
Sample $(k,i)\in\mathbb{N}\times\{1,2\}$ according to the discrete distribution
proportional to $\{w_{k,i}\}$.

\item \emph{Sampling within the band:}
Conditional on $(k,i)$, the density is uniform with respect to volume on the interval
$\|x\|\in[a\Delta,b\Delta]$, where
\[
[a,b]=
\begin{cases}
[k,k+\gamma], & i=1,\\
[k+\gamma,k+1], & i=2.
\end{cases}
\]
Since $V(r)\propto r^n$, the conditional CDF of $R=\|X\|$ is
\[
\Pr(R\le r)
=
\frac{r^n-(a\Delta)^n}{(b\Delta)^n-(a\Delta)^n},
\]
for $r\in[a\Delta,b\Delta]$. Using inverse transform sampling, we obtain 
\[ R=\Delta\big(I(b^n-a^n)+a^n\big)^{1/n},\]
where $I\sim\mathrm{Unif}[0,1]$. A direction $U$ is sampled by drawing $Y$ uniformly from the unit ball $\mathbb B$ of $\|\cdot\|$ and setting $U=\frac{Y}{\|Y\|}.$ The sample is then formed as $X = RU$. 

For the $\ell_1$-norm, this direction can be generated by drawing
$G_1,\dots,G_n\stackrel{\mathrm{iid}}{\sim}\mathrm{Exp}(1)$, setting
$A_i=G_i/\sum_jG_j$, and assigning independent random signs
$S_i\in\{\pm1\}$. The vector $(S_1A_1,\dots,S_nA_n)$ is uniformly distributed on the $\ell_1$-unit sphere. For the $\ell_2$-norm, one may draw $G\sim \mathcal N(0,I_n)$ and set $U=G/\|G\|_2$, which is uniform on the Euclidean unit sphere.

More generally, let $\mathcal K\subseteq\mathbb R^n$ be a symmetric convex body containing the origin in its interior, and let $\|\cdot\|_{\mathcal K}$ be its Minkowski functional. Since $\mathcal K$ is the unit ball of this norm, draw $Y$ uniformly from $\mathcal K$ and set $U=\frac{Y}{\|Y\|_{\mathcal K}}.$
Then $U$ lies on the $\|\cdot\|_{\mathcal K}$-unit sphere and has the
directional distribution induced by uniform sampling from $\mathcal K$. 
Drawing $R$ independently as above and returning $X=RU$ produces the staircase sample. The $\ell_1$- and $\ell_2$ constructions are special cases of this procedure. 
\end{itemize}

\added{The same sampler can also be used for non-isotropic norms, such
as the weighted $\ell_p$ norm
$\|x\|_{w,p}\coloneqq(\sum_{i=1}^n w_i|x_i|^p)^{1/p}$, where
$p\geq 1$ and $w_i>0$. To handle this norm, define
$A_w\coloneqq\operatorname{diag}(w_1^{1/p},\ldots,w_n^{1/p})$.
Since $\|x\|_{w,p}=\|A_wx\|_p$, the weighted unit ball is
$A_w^{-1}\mathbb B_p$. Hence, if $Y$ is sampled from the staircase
distribution corresponding to the standard $\ell_p$-norm with
parameters $(\eps,\Delta,\gamma)$, then $X=A_w^{-1}Y$ has the staircase
distribution corresponding to $\|\cdot\|_{w,p}$ with the same
parameters. Thus, weighted $\ell_p$ staircase noise can be generated
using the standard $\ell_p$ sampler followed by a diagonal rescaling.
Moreover, any cost of the form
$\Phi(x)=\varphi(\|x\|_{w,p})$ with $\varphi$ nondecreasing is
norm-monotone with respect to the weighted $\ell_p$ norm and therefore
falls within the scope of Theorem~\ref{thm:conjecture_resolution}.
As a particular example,
$\Phi(x)=\|x\|_{w,2}^2=\sum_i w_i x_i^2$ corresponds to the weighted mean-squared error.}

\added{While the band representation above is convenient for deriving the radial distribution, its direct implementation requires exact sampling from the countably supported distribution $\{w_{k,i}\}$; truncating this distribution would violate the pure $\eps$-DP guarantee. For exact implementation, we therefore use an equivalent sampling representation of the staircase distribution from~\cite{Staircase_General}. Let $T$ be
supported on $\{k+\gamma:k\in\mathbb Z_{\geq0}\}$ with $\Pr(T=k+\gamma)\propto(k+\gamma)^n e^{-k\eps},$
and let $Y$ be independent and uniformly distributed on the unit ball
$\mathbb B$ of $\|\cdot\|$. Kulesza et
al.~\cite{Staircase_General} refer to the distribution of $T$ as the
discrete Gamma distribution and give a rejection-sampling
algorithm for generating such $T$ in~\cite[Algorithm~1]{Staircase_General}.
The random variable $X=\Delta T Y$ has exactly the staircase density
$f_\gamma$. To see this, notice that conditional on $T=k+\gamma$, $X$ is uniform
on $(k+\gamma)\Delta\mathbb B$, whose volume is proportional
to $(k+\gamma)^n$. The factor $(k+\gamma)^n$ in the probability of
$T=k+\gamma$ therefore cancels with the volume of this ball, thus its
contribution to the density at $x$ is proportional to $e^{-k\eps}$
whenever $\|x\|\leq(k+\gamma)\Delta$. Summing over all such $k$ gives
$$\sum_{k:\,(k+\gamma)\Delta\geq\|x\|}e^{-k\eps}.$$
This geometric sum is constant on each of the two bands $[k\Delta,(k+\gamma)\Delta)$ and $[(k+\gamma)\Delta,(k+1)\Delta)$, with the density on the latter lower by a factor $e^{-\eps}$. Thus, this exact sampling representation reproduces the band representation above while avoiding any truncation of the infinite support.}

\begin{algorithm}[H]
\caption{Sampler of the staircase distribution}
\label{Alg:sampler}
\begin{algorithmic}[1]
\Require Parameters $(\varepsilon,\Delta,n,\gamma)$, norm $\|\cdot\|$, and a uniform sampler for its unit ball
\State Sample $T$ from the distribution
$\Pr(T=k+\gamma)\propto(k+\gamma)^n e^{-k\eps}$,
$k\in\mathbb Z_{\geq0}$, using the rejection sampler of
\cite[Algorithm~1]{Staircase_General}
\State Draw $Y$ uniformly from the $\|\cdot\|$-unit ball
\State \Return $X=\Delta T Y$
\end{algorithmic}
\end{algorithm}

\deleted{Since the weights decay exponentially in $k$, their total mass is negligible beyond a sufficiently large threshold $K_{\max}$. For
implementation, we therefore truncate the weights at $k\leq K_{\max}$
and renormalize them, choosing $K_{\max}$ so that the omitted tail mass
is below $10^{-12}$. Repeated i.i.d. samples from the staircase mechanism are obtained by calling Algorithm~\ref{Alg:sampler} independently.}
\added{It should be noted that Algorithm~\ref{Alg:sampler} assumes an exact uniform sampler for the
unit ball. If this subroutine is implemented only approximately, pure
$\eps$-DP is not automatic and the resulting sampling error must be
accounted for separately in the privacy guarantee.}

\begin{figure}
\centering
    \includegraphics[scale=0.11]{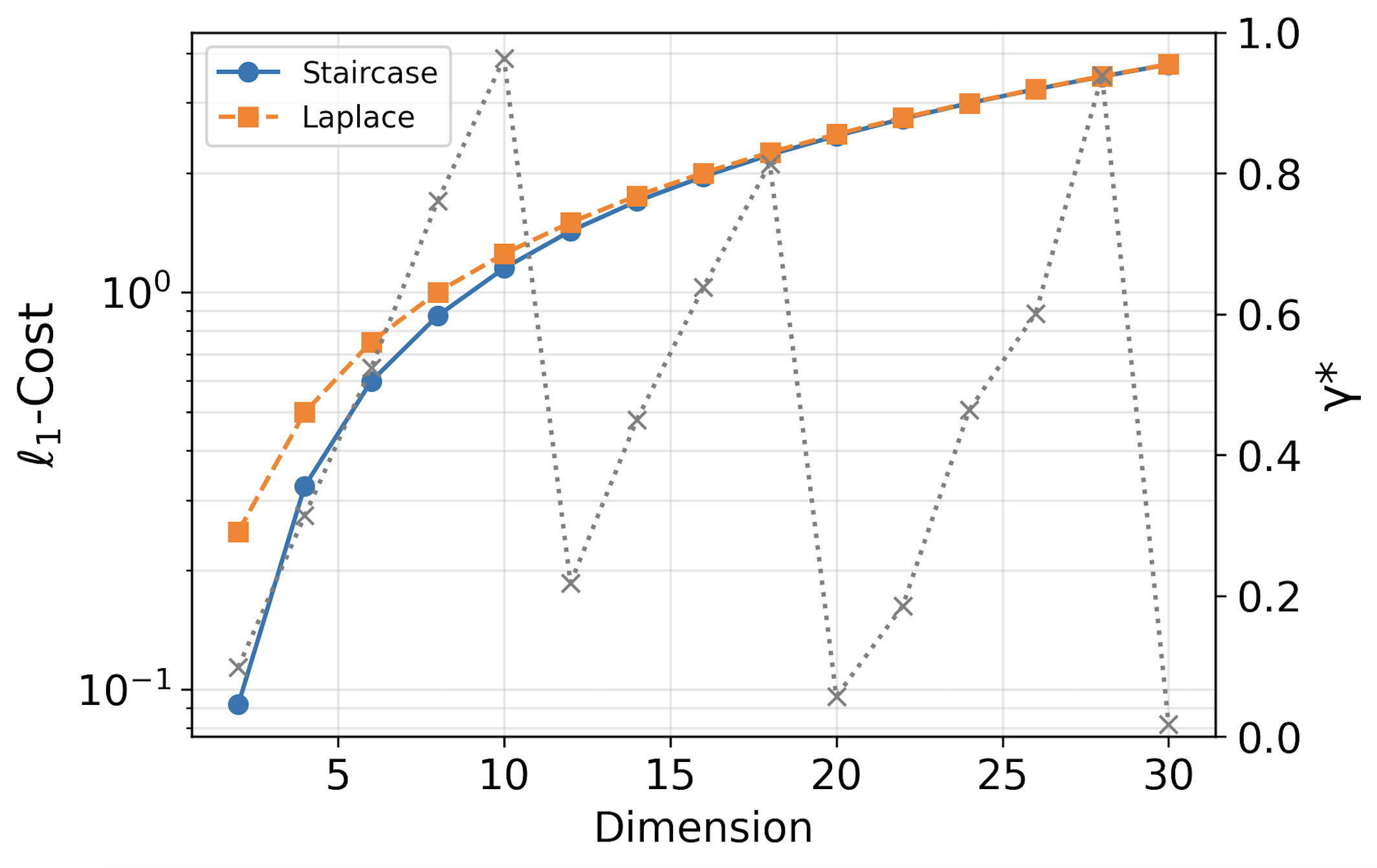}~~~~ \includegraphics[scale=0.11]{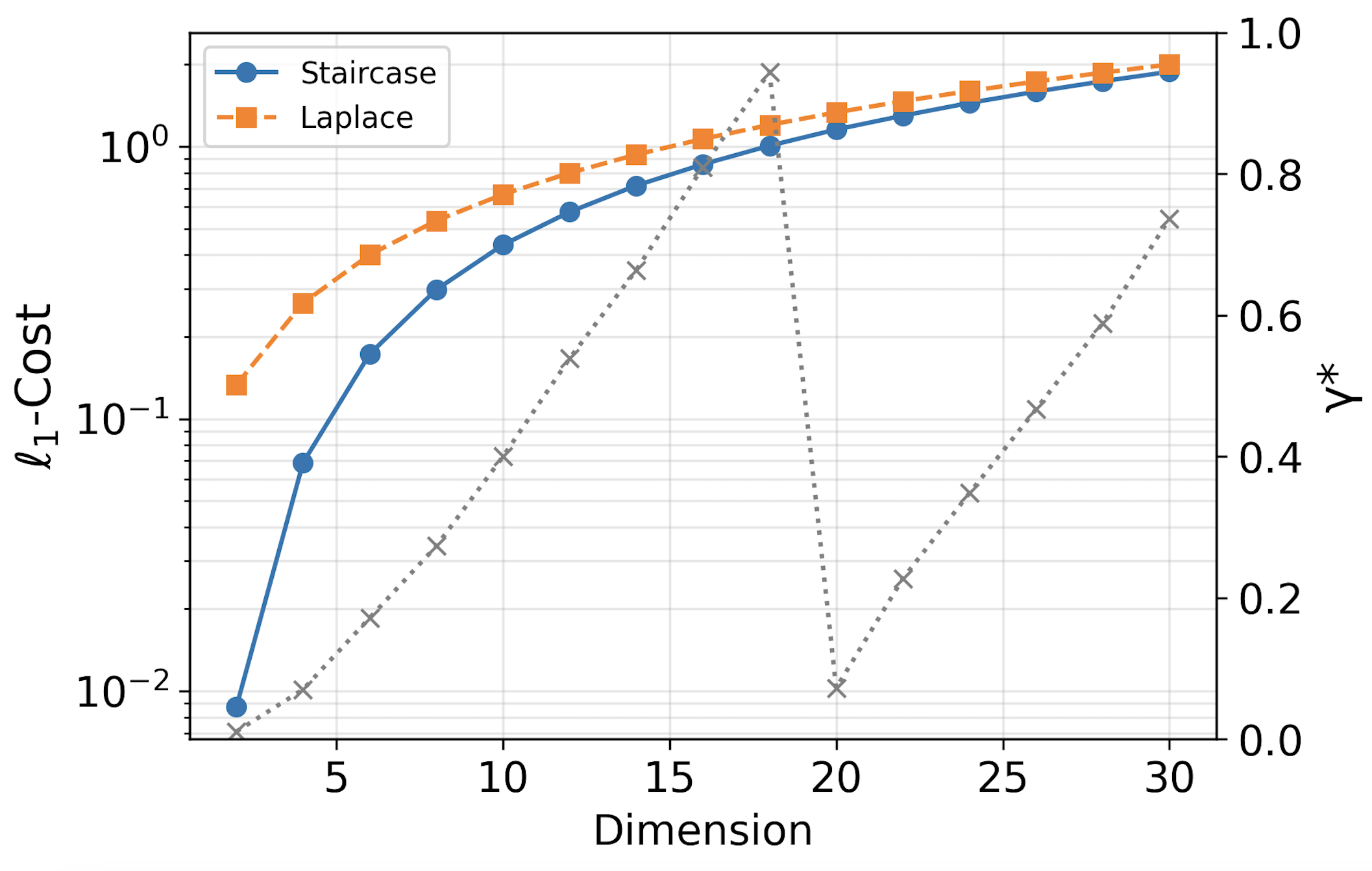}
    \caption{The utility-privacy tradeoff for staircase mechanism and Laplace mechanism for $\eps = 8$ (left) and $\eps = 15$ (right) with $\|\cdot\|_1$-sensitivity one, where the utility is measured via the cost function $\Phi(x) = \|x\|_1$. The gray dotted curve shows the optimal $\gamma^\ast$, the solution of the optimization problem \eqref{eq:Gamma^*}.}
    \label{fig:Delta1_epsilon}
\end{figure}
\subsection{Numerical Illustrations}
We now use the radial formula~\eqref{eq:radial-cost} to compare the
optimal staircase mechanism with standard pure-DP additive mechanisms.
For the moment costs considered below, we use the explicit
series~\eqref{eq:staircase-moment}. We evaluate this deterministic
objective for $\gamma\in[0,1]$ and approximate $\gamma^\ast$ by
one-dimensional grid search over a uniform grid on $[0,1]$ with spacing
$10^{-4}$. \added{For the infinite series
in~\eqref{eq:staircase-moment}, let $S_m^{(K)}(\gamma)$ denote the
partial sum over $0\leq k\leq K$ and set $q=e^{-\eps}$. Uniformly over
$\gamma\in[0,1]$, we have 
$$
0\leq S_m(\gamma)-S_m^{(K)}(\gamma)
\leq
\frac{m q^{K+1}(K+2)^{m-1}}
{1-q\big(\frac{K+3}{K+2}\big)^{m-1}},
$$
for $K$ large enough that the denominator is positive. We apply this bound to both
$S_n(\gamma)$ and $S_{n+p}(\gamma)$ to obtain upper and lower bounds
on the ratio in~\eqref{eq:staircase-moment}, and increase $K$ until
the resulting interval has width below $10^{-12}$.}

Figures~\ref{fig:Delta1} and~\ref{fig:Delta1_epsilon} illustrate the utility gap between the optimal staircase and Laplace mechanisms under $\|\cdot\|_1$-sensitivity $\Delta=1$. As guaranteed by our optimality theorem, the staircase cost is no larger than the Laplace cost. The figures quantify the size of this theoretically established gap across privacy levels and dimensions.

\begin{figure}
    \centering
    \includegraphics[width=0.5\linewidth]{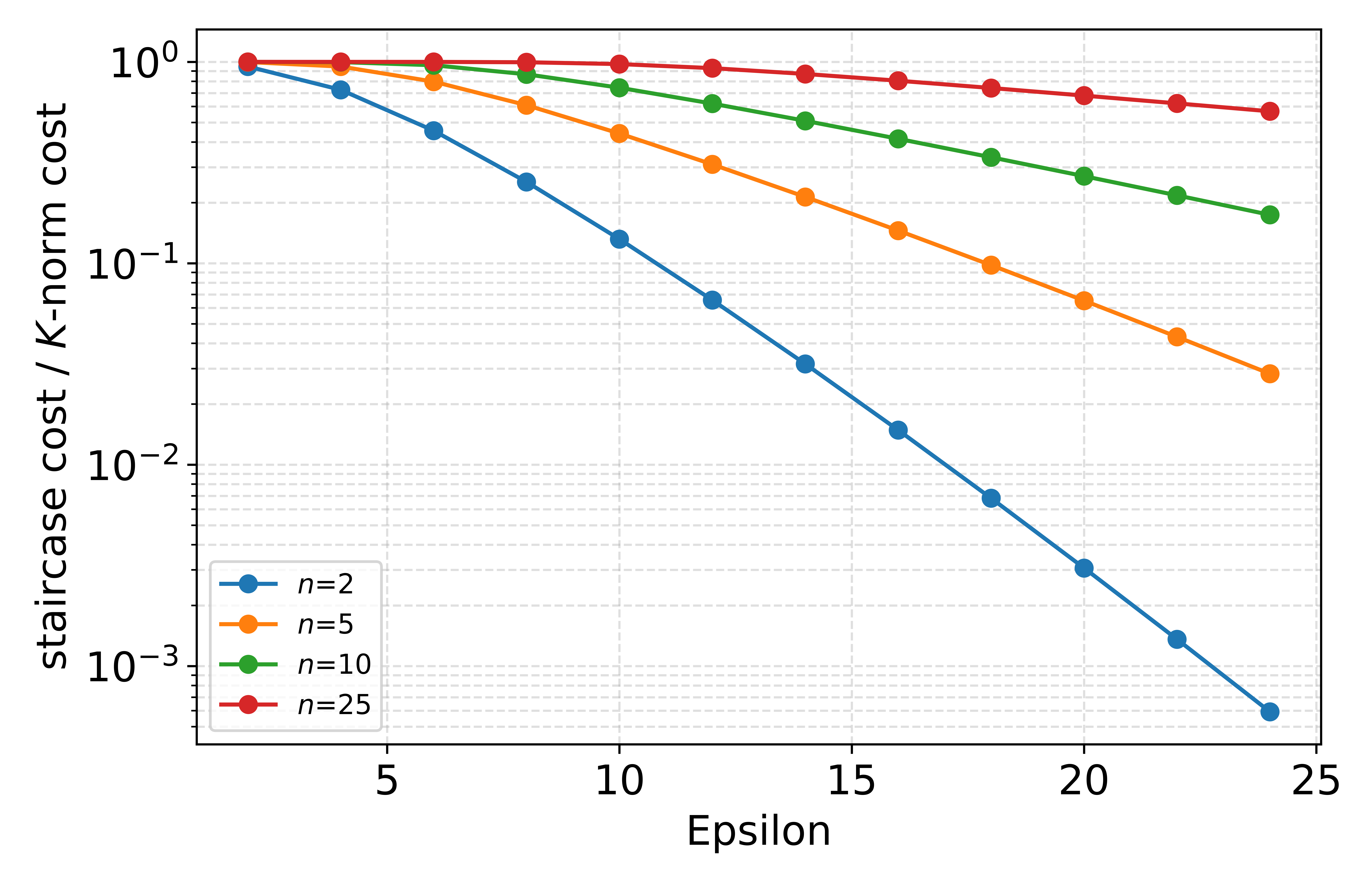}
    \caption{Multiplicative improvement of the staircase mechanism over the corresponding $\mathcal K$-norm mechanism. The privacy geometry is $\ell_2$-sensitivity and the cost is $\mathbb E\|X\|_2^2$, corresponding to the noise power/MSE of the additive mechanism. Values below one indicate lower cost for the staircase mechanism.}
    \label{fig:knorm}
\end{figure}

We also compare the staircase mechanism with the corresponding $\mathcal K$-norm mechanism \cite{HardtGeometry,Awan03042021}. Recall that a $\mathcal K$-norm mechanism is an additive mechanism whose noise density is proportional to $\exp(-\varepsilon\|x\|_{\mathcal K}),$
where $\|\cdot\|_{\mathcal K}$ is the Minkowski functional induced by a convex sensitivity body $\mathcal K$, namely
$\|x\|_{\mathcal K}\coloneqq \inf\{r\ge 0:x\in r\mathcal K\}.$
In our norm-sensitivity formulation, if $\mathcal K=\Delta\mathbb B$ with respect to a norm $\|\cdot\|$, then $\|x\|_{\mathcal K}=\|x\|/\Delta$, where $\Delta$ is the $\|\cdot\|$-sensitivity. Hence the corresponding $\mathcal K$-norm density is proportional to
$\exp(-\varepsilon\|x\|/\Delta).$ In the Euclidean setting, where $\|\cdot\|=\|\cdot\|_2$, this becomes $\exp(-\varepsilon\|x\|_2/\Delta)$. Passing to polar coordinates, the radial variable $R=\|X\|_2$ has density proportional to $r^{n-1}\exp(-\varepsilon r/\Delta)$. Therefore,
$$\mathbb E_{\mathcal K\text{-norm}}\left[\|X\|_2^2\right]=\mathbb E[R^2]=\Big(\frac{\Delta}{\varepsilon}\Big)^2
\frac{\Gamma(n+2)}{\Gamma(n)}=\frac{n(n+1)\Delta^2}{\varepsilon^2}.$$
For the staircase mechanism, we compute $\mathbb E_{f_{\gamma}}[\|X\|_2^2]$ using the same radial-band evaluation described above and optimize over $\gamma\in [0,1]$. 
Figure~\ref{fig:knorm} reports the ratio between the cost of the  resulting staircase mechanism $M_{\gamma^\ast}$ and that of the corresponding $\mathcal K$-norm mechanism. This ratio is at most one
and is smallest for moderate-to-large $\eps$ and lower dimensions,
where the staircase mechanism provides the largest improvement. As
$n$ increases or $\eps$ decreases, the ratio approaches one and the
improvement diminishes. 
\added{The same comparison also applies to the weighted $\ell_2$-norm.
Indeed, under the change of variables $y=A_wx$ described above, the
weighted staircase and corresponding $\mathcal K$-norm mechanisms
reduce to their Euclidean counterparts, while
$\|x\|_{w,2}^2=\|y\|_2^2$. Thus, for fixed $(n,\eps,\Delta)$, the
staircase-to-$\mathcal K$-norm cost ratio under weighted $\ell_2$-sensitivity and weighted mean-squared error is exactly the one reported
in Fig.~\ref{fig:knorm}.}

These numerical results provide a quantitative illustration of the gains predicted by Theorem~\ref{thm:conjecture_resolution}. In particular, they show that the staircase mechanism can yield non-trivial improvements over standard pure-DP additive mechanisms, including the Laplace mechanism and the $\mathcal K$-norm mechanism, in some parameter regimes.

\section{Discussion and Limitations}

\added{In this work, we show that, for any prescribed norm $\|\cdot\|$ and any
norm-monotone cost $\Phi(x)=\varphi(\|x\|)$, an optimal noise distribution in
$\mathcal M_n^{\|\cdot\|}(\Delta,\eps)$ can be chosen from the
staircase family. In fact, our framework yields a stronger structural result:
if $X$ is the noise variable of any feasible query-independent additive
mechanism and $X^\ast$ is its rearrangement, then replacing $X$ by
$X^\ast$ preserves $\eps$-DP and
$\Pr(\|X^\ast\|>t)\leq\Pr(\|X\|>t)$ for every $t\geq0$. Consequently,
$\mathbb E[\varphi(\|X^\ast\|)]\leq
\mathbb E[\varphi(\|X\|)]$ for every nondecreasing $\varphi$, thus the
rearrangement simultaneously improves all norm-monotone cost functions. This
reduces the optimal noise-design problem to a one-dimensional radial
optimization, whose extreme points are the staircase distributions.
Hence, for every norm-monotone cost function, an optimizer lies in the staircase
family, although the optimal $\gamma^\ast$ may depend on $\Phi$.}

\added{Our result is complementary to that of Kulesza, Suresh, and
Wang~\cite{Staircase_General}. Their query-specific formulation exploits
the full sensitivity space $\mathcal K$ and, when $\mathcal K$ is
convex, proves optimality of the generalized staircase mechanism for
the matching first-moment cost $\mathbb E[\|X\|_{\mathcal K}]$. In the
norm-ball setting $\mathcal K=\Delta\bB$, their staircase family
coincides with ours and their result recovers the special case
$\Phi(x)=\|x\|$. Our formulation instead follows the standard
sensitivity-calibrated model in which a mechanism is specified by a
prescribed norm and its scalar sensitivity $\Delta$, without requiring
the full query-specific geometry of $\mathcal K$. Within this
query-independent model, our result extends staircase optimality from
the first-moment norm loss to every norm-monotone cost.}

\added{This distinction also identifies a principal limitation of our
framework: the privacy and utility geometries must be aligned through
the same prescribed norm. For example, suppose
$\operatorname{conv}(\mathcal K)=\bB_\infty$ while utility is measured
by $\Phi(x)=\|x\|_2^2$. If privacy is calibrated using
$\|\cdot\|_\infty$, then $\Phi$ is not norm-monotone with respect to
that norm and our theorem does not apply. Alternatively, using the
$\ell_2$-norm aligns the utility with our framework, but the
corresponding sensitivity is $\Delta=\sqrt n$, thus our theorem applies
over $\mathcal M_n^{\|\cdot\|_2}(\sqrt n,\eps)$. Since
$\bB_\infty\subsetneq\sqrt n\,\bB_2$, this norm-ball formulation
imposes stronger privacy constraints than those induced by the
$\ell_\infty$ sensitivity body and may therefore incur additional
utility loss. More generally, the query-independent norm-ball
formulation can be conservative when the exact sensitivity space is a
proper subset of the enclosing norm ball. Characterizing optimal
additive mechanisms when the sensitivity and utility geometries are
mismatched is a natural direction for future work.}

\bibliographystyle{IEEEtran}
\bibliography{reference}

\appendices

\section{Proof of Theorem~\ref{Thm:DPequivalentEnlarged}} \label{sec: ep DP is Lipschitz Condition}
Theorem~\ref{Thm:DPequivalentEnlarged} provides a geometric characterization of
$\varepsilon$-DP for additive mechanisms in terms of a
log-Lipschitz property of the noise density with respect to the metric
\[
d_{\varepsilon,\Delta}(x,y)
\coloneqq
\varepsilon\Big\lceil\tfrac{\|x-y\|}{\Delta}\Big\rceil .
\]
The proof proceeds in two steps. First, we show that any noise distribution
satisfying the DP shift condition must be absolutely continuous with respect to Lebesgue measure. Second, we establish the equivalence between log-Lipschitzness and a level-set enlargement property.

\subsection{Absolute continuity of DP noise distributions}

We begin by establishing a general absolute continuity result.

\begin{theorem}\label{thm:ExistenceDensity}
Let $C>0$ and $\delta>0$, and let $\mu$ be a probability measure on $\mathbb{R}^n$
satisfying
\begin{align}\label{eq:ConditionExistenceDensity} 
\mu(A) &\leq C~ \mu(A+x), \ \mbox{ or equivalently,} \\
{\mu(A + x)} &\leq C \mu(A) \nonumber
\end{align}
for all Borel sets $A\subset\mathbb{R}^n$ and all $x$ with $\|x\|\le\delta$.
Then $\mu$ is absolutely continuous with respect to Lebesgue measure.
\end{theorem}

{Note, since norms on $\mathbb{R}^n$ are comparable up to constants the result is independent of choice of norm; we therefore take $\|x\| \coloneqq \|x\|_\infty = \max_i |x_i|$ for convenience.  The proof of the equivalence between the two inequalities follows immediately by considering $\tilde{A} = A +x$ and $\tilde{x} = -x$, what remains} relies on the following auxiliary lemma.

\begin{lemma}\label{lem:ExistenceDensityHyperplanes}
Under the assumptions of Theorem~\ref{thm:ExistenceDensity},
$\mu(H)=0$ for every affine hyperplane $H\subset\mathbb{R}^n$.
\end{lemma}

\begin{proof}
Any affine hyperplane can be written as $H=u+H_v$, where
$H_v=\{x\in\mathbb{R}^n:\langle x,v\rangle=0\}$ for some $v$ with $\|v\| = 1$.
The translates $\{tv+H_v\}_{t\in\mathbb{R}}$ are pairwise disjoint.
For $|t|\le\delta$, the assumption implies
\[
\mu(tv+H_v)\ge C^{-1}\mu(H_v).
\]
Summing over $t\in[-\delta,\delta]\cap\mathbb{Q}$ and using that $\mu$ is a
probability measure forces $\mu(H_v)=0$, and hence $\mu(H)=0$.
\end{proof}

We also need  the following property of open sets. Following the notation by Folland \cite{folland1999real}, we let $\mathcal{Q}_{k}$ be the collection of (closed) cubes whose side length is $2^{-k}$ and whose vertices are in the lattice $(2^{-k}\mathbb{Z})^{n}$. The \textit{dyadic} cubes are the cubes that belong to $\mathcal{Q}_{k}$ for some $k\in\mathbb{N}$.

\begin{lemma}[Lemma~2.43, \cite{folland1999real}]
\label{lem:Folland}
If $A\subset\mathbb{R}^{n}$ is open, then $\displaystyle A = \bigcup_{k=1}^\infty \mathcal{V}(A,k)$ where
\begin{equation*}
    \mathcal{V}(A,k) \coloneqq \bigcup \{ Q \in \mathcal{Q}_k : Q \subset A \}.
\end{equation*}
Moreover, $A$ is a countable union of dyadic cubes with disjoint interiors.
\end{lemma}

Now we are in a position to prove Theorem~\ref{thm:ExistenceDensity}.

\begin{proof}[\textbf{{Proof of Theorem}~\ref{thm:ExistenceDensity}}] 
We want to prove that $\mu(A) = 0$ for every Borel set $A\subseteq \mathbb{R}^{n}$ with $\lvert A \rvert = 0$.  By scaling, (taking $\mu_\lambda(A) = \mu( \lambda A)$, for $\lambda > 0$) we may assume that $\delta = 1.$

Let $I \coloneqq [0,1]^{n}$. Observe that, for any Borel set $A\subseteq \mathbb{R}^{n}$, we have
\begin{align*}
    \mu(A) &= \mu\bigg(\bigcup_{v\in\mathbb{Z}^{n}} A \cap (v+I)\bigg)\\
    &\leq \sum_{v\in\mathbb{Z}^{n}} \mu(A \cap (v+I))\\
    &\leq \sum_{v\in\mathbb{Z}^{n}} C^{  \lceil \lVert -v \rVert \rceil} \mu((A-v) \cap I),
\end{align*}
where the last inequality follows from inducting on \eqref{eq:ConditionExistenceDensity}. By the translation invariance of the Lebesgue measure $|(A-v) \cap I | \leq |A-v| = |A| = 0$, thus if the result holds for sets contained in $I$, it holds in general, as the right hand sum will be zero.   Note  the above inequality forces $\mu(I) > 0$, otherwise $\mu = 0$. In order to establish the theorem, it is enough to prove that there exists $c>0$ such that, for every Borel set $E \subset \mathbb{R}^{n}$,
\begin{equation}
\label{eq:ExistenceDensityInq}
    \mu(E \cap I) \leq c \  \lvert E \cap I \rvert.
\end{equation}

We start proving \eqref{eq:ExistenceDensityInq} for dyadic cubes $Q \subset I$. For $k\in\mathbb{N}$, let $\Lambda_{k} \coloneqq \{0,1,\ldots,2^{k}-1\}/2^{k}$. By Lemma~\ref{lem:ExistenceDensityHyperplanes} 
\begin{equation}
\label{eq:ExistenceDensityVkn}
    \mu(I) = \sum_{v\in \Lambda_{k}^{n}} \mu\Big(v+\frac{1}{2^{k}}I\Big).
\end{equation}
For $v_{0},v\in \Lambda_{k}^{n} \subseteq I$, $\|v_0 - v \| \leq 1 = \delta$, so that by
\eqref{eq:ConditionExistenceDensity}, 
\begin{equation*}
   \mu\Big(v_0 + \frac{1}{2^{k}} I\Big) \leq {C} \mu\left((v- v_0) + v_0  + \frac 1 {2^k} I\right) = C \mu \Big(v + \frac{1}{2^{k}} I\Big).
\end{equation*}

Inserting this inequality in \eqref{eq:ExistenceDensityVkn}, and using that the number of elements in $\Lambda_k^n$ is  $2^{kn} = | v_0 + \frac{1}{2^k} I|^{-1}$ we obtain 
\begin{align*}
            \mu(I) 
        \geq \ \frac{ \mu \left(v_0 + \frac{1}{2^k} I \right) }{C \left|v_0 + \frac{1}{2^k} I \right|}. 
\end{align*}
Writing 
$
    c \coloneqq C \mu(I),
$
we have 
\[
    { \mu(v_0 + \tfrac{1}{2^k} I) \leq c |v_0 + \tfrac{1}{2^k} I |}.
\]
Given that $k$ was general and $v_0$ was an arbitrary vertex, we have that for dyadic cubes contained in $I$,
\begin{align} \label{eq: dyadic cube inequality}
    \mu(Q) \leq c |Q|.
\end{align}
Let $A \subseteq I$ be an open set. Lemma~\ref{lem:Folland} establishes that $\displaystyle A = \bigcup_{j\geq1} Q_{j}$ where $\{Q_{j}\}_{j\geq1}$ is a sequence of dyadic cubes with disjoint interiors. Since neither $\mu$ nor the Lebesgue measure assign positive measure to hyperplanes and \eqref{eq: dyadic cube inequality}, we can write 
\begin{equation*}
    \mu(A) = \sum_{j\geq1} \mu (Q_{j}) \leq c \sum_{j \geq 1} |Q_j| = c |A|.
\end{equation*}
To extend to general Borel $A$, fix $\eta > 0$  and first suppose that $A \subseteq [\eta, 1 - \eta]^n$, considering an open set $\mathcal{O}$ such that $ A \subseteq \mathcal{O} \subseteq I$, we have
\[
    \mu(A) \leq \mu(\mathcal{O}) \leq c  \ |\mathcal{O}|.
\]
Taking the infimum\footnote{More precisely the regularity of the Lebesgue measure gives $|A| = \inf \{ |U|: A \subseteq U \mbox{ open } \} $.  Since $A \subseteq [\eta, 1- \eta]^n $, for a given $U$ containing $A$, the set $\mathcal{O} = (\tfrac{\eta}{2}, 1 - \tfrac{\eta}{2})^n \ \cap U$ will still contain $A$ and thus we can restrict our infimum to open sets contained in $I$.} over all such $\mathcal{O}$ and using the regularity of the Lebesgue measure we have
\[
    \mu(A) \leq c \ |A|.
\]
Finally, using again that neither $\mu$ nor the Lebesgue measure give mass to hyperplanes and then the continuity of measure
\begin{align*}
    \mu(A) 
        &= 
            \mu( A \cap (0,1)^n )
                \\
        &= 
            \mu \left( A \cap \left( \bigcup_{k=1}^\infty [\tfrac 1 k , 1 - \tfrac 1k]^n \right) \right)
                \\
        &=
            \lim_{k \to \infty} \mu(A \cap [\tfrac 1k , 1 - \tfrac 1k]^n )
                \\
        &\leq
            \lim_{k \to \infty} c |A \cap [\tfrac 1k , 1- \tfrac 1k]^n |
                \\
        &=
            c |A|.
\end{align*}
\end{proof}

\subsection{Lipschitz functions and level-set enlargements}

We next record a general equivalence between Lipschitz continuity and level-set
containment.

\begin{lemma}\label{lem: set-theoretic 1 lip}
Let $(E,d)$ be a metric space and let $g:E\to\mathbb{R}$.
Then $g$ is $1$-Lipschitz if and only if
\[
\{g>\lambda\}_{h} \subseteq \{g>\lambda-h\}
\]
for all $\lambda\in\mathbb{R}$ and all $h>0$.
\end{lemma}
\begin{proof}
    Suppose $g$ satisfies the set-theoretic containment. Take $x,y \in E$, $\delta >0$, and $h = d(x,y) + \delta$.  Then, we have  
\[
    y \in \{ g > g(x) - \delta \}_{d(x,y) + \delta} \subseteq \{ g > g(x) - 2\delta - d(x,y) \}.
\]
Thus,
\[
    g(x) - g(y) < d(x,y) + 2 \delta.
\]
Reversing the roles of $x,y$ we have
\[
    |g(x) - g(y)| < d(x,y) + 2 \delta.
\]
Taking $\delta$ to zero, shows that $g$ is $1$-Lipschitz.
Conversely, assume that $g$ is $1$-Lipschitz and consider $x \in \{ g > \lambda \}_h$ for $\lambda \in \mathbb{R}$, by definition of an $h$-enlargement there exists $y \in E$ such that $g(y) > \lambda$ and $d(x,y) < h$.  Using the lower bound on $g(y)$ and the fact that $g$ is $1$-Lipschitz,
\[
    \lambda -g(x) < g(y) - g(x) < d(x,y) < h.
\]
Rearranging the inequality shows that $g(x) > \lambda - h$ or $x \in \{ g > \lambda - h \}$.
\end{proof}

\subsection{Proof of Theorem~\ref{Thm:DPequivalentEnlarged}} 
\begin{proof}
We prove two directions.
\begin{description}
    \item[DP implies level-set inclusion ($\Rightarrow$):]   Assume that $M_X(D)=q(D)+X$ is $\varepsilon$-DP for every query $q$ with
$\|\cdot\|$-sensitivity $\Delta$.
Let $\mu$ denote the law of $X$.
For any $u\in\mathbb{R}^n$ with $\|u\|\le\Delta$ and any Borel set $A$,
the DP inequality implies
\[
\mu(A)\le e^\varepsilon \mu(A+u).
\]
Thus $\mu$ satisfies the hypothesis of Theorem~\ref{thm:ExistenceDensity} with
$C=e^\varepsilon$ and $\delta=\Delta$, and hence $X$ admits a density $f$.

Fix $x,y\in\mathbb{R}^n$ with $\|x-y\|\le\Delta$.
Applying the above inequality to the sets $A=y+\mathbb{B}_t$ and using translation
invariance of the sets gives
\[
\mu(x+\mathbb{B}_t)\le e^\varepsilon \mu(y+\mathbb{B}_t)
\qquad \forall\,t>0.
\]

{ Thus, $f_X(x) \coloneqq \limsup_{t \to 0} \frac{\mu(x + \mathbb{B}_t)}{|\mathbb{B}_t|}$ satisfies}
\[
    f_X(x) \leq e^{\varepsilon} f_X(y)
\]
However, Lebesgue’s differentiation
theorem yields
\[
f_X = f,
\]
{almost surely.}
Thus, we consider $f_X$ {\it the} density of $X$ and observe that it satisfies

\[
\log f_X(x)-\log f_X(y)\le \varepsilon
\qquad \text{whenever } \|x-y\|\le\Delta.
\]

Now let $x,y\in\mathbb{R}^n$ be arbitrary and define
$k\coloneqq\lceil \|x-y\|/\Delta\rceil$.
There exist points $x=x_0,x_1,\dots,x_k=y$ such that
$\|x_i-x_{i-1}\|\le\Delta$ for all $i$.
Summing the above inequality along this chain gives
\[
\log f_X(x)-\log f_X(y)\le k\varepsilon
= d_{\varepsilon,\Delta}(x,y).
\]
Repeating the argument with $x$ and $y$ interchanged yields
\[
|\log f_X(x)-\log f_X(y)|\le d_{\varepsilon,\Delta}(x,y),
\]
so $\log f_X$ is $1$-Lipschitz with respect to the metric $d_{\varepsilon,\Delta}$.

\deleted{Applying Lemma~\ref{lem: set-theoretic 1 lip} to the function
$g=\log f_X$ on the metric space $(\mathbb{R}^n,d_{\varepsilon,\Delta})$,
we obtain that for all $\lambda \in \mathbb{R}$ and $h>0$, that 
$\{\log f_X>\log\lambda\}_h\subseteq\{\log f_X>\log\lambda-h\}.$ Exponentiating both sides yields \(\{f_X>\lambda\}_h\subseteq\{f_X>\lambda e^{-h}\},\) which is exactly \eqref{eq:dpenlargement}.}
\added{Applying Lemma~\ref{lem: set-theoretic 1 lip} to the function
$g=\log f_X$ on the metric space
$(\mathbb{R}^n,d_{\varepsilon,\Delta})$, we obtain, for every
$\alpha\in\mathbb{R}$ and $h>0$,
\[\{\log f_X>\alpha\}_h\subseteq\{\log f_X>\alpha-h\}.\]
Setting $\lambda=e^\alpha>0$ gives
\[\{f_X>\lambda\}_h\subseteq\{f_X>\lambda e^{-h}\},\]
which is exactly~\eqref{eq:dpenlargement}.}
\item[Level-set inclusion implies DP ($\Leftarrow$):] Conversely, suppose that the density $f_X$ satisfies the level-set inclusion
\eqref{eq:dpenlargement}.
By Lemma~\ref{lem: set-theoretic 1 lip}, this is equivalent to the function
$\log f_X$ being $1$-Lipschitz with respect to $d_{\varepsilon,\Delta}$.
In particular, for any $z\in\mathbb{R}^n$ and any $u\in\mathbb{R}^n$ with
$\|u\|\le\Delta$, we have
$d_{\varepsilon,\Delta}(z,z+u)\le\varepsilon$, and hence
\[
f_X(z+u)\le e^\varepsilon f_X(z).
\]

Let $q$ be any query with $\|\cdot\|$-sensitivity $\Delta$, and let $D\sim D'$ be
neighboring datasets.
Writing $u\coloneqq q(D')-q(D)$, we have $\|u\|\le\Delta$, and therefore
\begin{align*}
\Pr(M_X(D)\in A)
&=
\int_{A-q(D)} f_X(z)\,\mathrm{d}z
=
\int_{A-q(D')} f_X(z+u)\,\mathrm{d}z \\
&\le
e^\varepsilon \int_{A-q(D')} f_X(z)\,\mathrm{d}z
=
e^\varepsilon \Pr(M_X(D')\in A),
\end{align*}
for all measurable $A$.
Thus $M_X$ is $\varepsilon$-DP.
\end{description}
This completes the proof.
\end{proof}

\section{Proof of Theorem~\ref{Prop:NormRearrangementSmaller}}
\label{sec: Proof of Theorem 2}

We first establish several basic properties of radially symmetric decreasing
rearrangements.

\subsection{Proof of Proposition~\ref{Prop:PropertiesRearrangements}}
\begin{proof}
\begin{enumerate}[label=(\roman*)]

\item
By Definition~\ref{Def:ConvexRearrangement}, $A^\ast$ is a centered $\|\cdot\|$-ball
with Lebesgue measure equal to that of $A$. Since Lebesgue measure is homogeneous,
$|A^\ast|=|A|$.

\item
Assume that $x\in\{f^{\ast}>\lambda\}$. In particular, we have that
\begin{equation}
\label{eq:PropertiesRearrangemetsProofb1}
    f^{\ast}(x) = \int_{0}^{\infty} \mathbbm{1}_{\{f>t\}^{\ast}}(x) \,\mathrm{d}t > \lambda.
\end{equation}
Since $\{f>t_{1}\}^{\ast} \supseteq \{f>t_{2}\}^{\ast}$ for $t_{1}<t_{2}$, we have that
\begin{equation*}
    \mathbbm{1}_{\{f>t_{1}\}^{\ast}}(x) \geq \mathbbm{1}_{\{f>t_{2}\}^{\ast}}.
\end{equation*}
This inequality implies that there exists $t_{0}\geq0$ such that
\begin{equation*}
    \mathbbm{1}_{\{f>t\}^{\ast}}(x) = \begin{cases} 1 & t<t_{0},\\ 0 & t>t_{0}.\end{cases}
\end{equation*}
The previous expression and \eqref{eq:PropertiesRearrangemetsProofb1} imply that $\lambda<t_{0}$. We conclude that $\mathbbm{1}_{\{f>\lambda\}^{\ast}}(x)=1$ and that $\{f^{\ast}>\lambda\} \subset \{f>\lambda\}^{\ast}$. \added{Since the rearrangement of a set is an open norm ball, the super-level
sets of $f^\ast$ are open, and hence $f^\ast$ is lower
semicontinuous.}

By continuity of measure, we have that
\begin{align}
    \nonumber \{f>\lambda\}^{\ast} &= \bigg(\frac{\lvert \{f>\lambda\} \rvert}{\lvert \mathbb{B} \rvert}\bigg)^{1/n} \mathbb{B}\\
    \nonumber &= \lim_{\eta\to0} \bigg(\frac{\lvert \{f>\lambda+\eta\} \rvert}{\lvert \mathbb{B} \rvert}\bigg)^{1/n} \mathbb{B}\\
    \label{eq:PropertiesRearrangemetsProofb2} &= \lim_{\eta\to0} \{f>\lambda+\eta\}^{\ast}.
\end{align}
Observe that if $x\in\{f>\lambda+\eta\}^{\ast}$, then $x\in\{f>t\}^{\ast}$ for all $t\leq\lambda+\eta$. In particular, we have that
\begin{equation*}
    f^{\ast}(x) = \int_{0}^{\infty} \mathbbm{1}_{\{f>t\}^{\ast}}(x) \,\mathrm{d}t \geq \lambda+\eta,
\end{equation*}
i.e., $x\in\{f^{\ast}\geq\lambda+\eta\}$. From this inclusion and \eqref{eq:PropertiesRearrangemetsProofb2}, we conclude that
\begin{align*}
    \{f>\lambda\}^{\ast} &\subseteq \lim_{\eta\to0} \{f^{\ast}\geq\lambda+\eta\}\\
    &= \{f^{\ast}>\lambda\}.
\end{align*}
The proof is now complete. 

\item
Recall that $A^{\ast} = s\mathbb{B}$ and $B^{\ast} = t\mathbb{B}$ with
\begin{equation*}
    s = \frac{\lvert A \rvert^{1/n}}{\lvert \mathbb{B} \rvert^{1/n}} \quad \text{and} \quad t = \frac{\lvert B \rvert^{1/n}}{\lvert \mathbb{B} \rvert^{1/n}}.
\end{equation*}
A straightforward manipulation shows that
\begin{align*}
    \lvert A^{\ast} \cap B^{\ast} \rvert &= \lvert \min\{s,t\} \mathbb{B} \rvert\\
    &= \min \{ \lvert A \rvert, \lvert B \rvert\}\\
    &\geq \lvert A \cap B \rvert.
\end{align*}
Therefore, we conclude that
\begin{align*}
    \lvert A^{\ast} \cap (B^{\ast})^c \rvert &= \lvert A^{\ast} \rvert - \lvert A^{\ast} \cap B^{\ast} \rvert\\
    &\leq \lvert A \rvert - \lvert A \cap B\rvert\\
    &= \lvert A \cap B^c \rvert.
\end{align*}

\item Previous item states that
\begin{equation}\label{eq:rearr_comp}
|A^\ast\cap (B^\ast)^c|\;\le\;|A\cap B^c|.
\end{equation}
Decompose $A^\ast$ as a disjoint union:
\[
A^\ast
=
(A^\ast\cap B^\ast)\ \cup\ (A^\ast\cap (B^\ast)^c),
\]
hence
\begin{equation}\label{eq:decomp_star_vi}
|A^\ast\cap B^\ast|
=
|A^\ast|-|A^\ast\cap (B^\ast)^c|.
\end{equation}
Similarly,
\begin{equation}\label{eq:decomp_vi}
|A\cap B|
=
|A|-|A\cap B^c|.
\end{equation}
Combining \eqref{eq:decomp_star_vi} with \eqref{eq:rearr_comp} gives
\[
|A^\ast\cap B^\ast|
=
|A^\ast|-|A^\ast\cap (B^\ast)^c|
\ge
|A^\ast|-|A\cap B^c|.
\]
Since $|A^\ast|=|A|$ (item (i)), we conclude
\[|A^\ast\cap B^\ast| \ge |A|-|A\cap B^c|=|A\cap B|,\]

\item This is a restatement of the Brunn-Minkowski inequality, see for example \cite[Thm.~4.1]{gardner2002brunn}, which states that for non-empty Borel measurable sets $A, B \subseteq \mathbb{R}^n$,
\[
    |A + B| \geq \left( |A|^{\frac 1 n} + |B|^{\frac 1 n } \right)^n.
\]By  Definition~\ref{Def:ConvexRearrangement}, it can be { directly} verified that
\begin{equation*}
    A^{\ast} + B^{\ast} = \frac{\lvert A \rvert^{1/n} + \lvert B \rvert^{1/n}}{\lvert \mathbb{B} \rvert^{1/n}} \mathbb{B},
\end{equation*}
which implies that $\lvert A^{\ast} + B^{\ast} \rvert = (\lvert A \rvert^{1/n} + \lvert B \rvert^{1/n})^{n}$. Hence, by the Brunn-Minkowski inequality,
\begin{equation*}
    \lvert A^{\ast} + B^{\ast} \rvert \leq \lvert A + B \rvert,
\end{equation*}
as required.

\item
If $\Psi:\mathbb{R}_+\to\mathbb{R}$ is strictly increasing and continuous, then
\[
\{\Psi(f)>\lambda\}
=
\{f>\Psi^{-1}(\lambda)\}.
\]
Applying part~(ii) yields
\[
\{(\Psi\circ f)^\ast>\lambda\}
=
\{f^\ast>\Psi^{-1}(\lambda)\}
=
\{\Psi(f^\ast)>\lambda\},
\]
which proves $\Psi\circ f^\ast=(\Psi\circ f)^\ast$.
\end{enumerate}
\end{proof}
\subsection{Stochastic domination}

Recall that for non-negative random variables $U,V$,
$$ U\prec V ~~ \iff ~~ \Pr(U>\lambda)\le\Pr(V>\lambda),$$
for all $\lambda\ge0.$
Equivalently,
\[U\prec V~~\iff~~\mathbb{E}[f(U)]\le\mathbb{E}[f(V)]\]
for all increasing $f$; see~\cite{shaked2007stochastic}.

We will use the following sufficient condition.

\begin{lemma}\label{Lemma:DensityStochasticDomination}
Let $U,V$ be non-negative random variables with densities $f,g$.
If there exists $t_0\ge0$ such that
$f(t)\ge g(t)$ for $t<t_0$ and $f(t)\le g(t)$ for $t>t_0$, then $U\prec V$.
\end{lemma}

\begin{proof}
For any increasing $\psi$,
\[ \mathbb{E}[\psi(V)-\psi(U)] = \mathbb{E}\left( \psi(V) - \psi(t_0) + \psi(t_0) -  \psi(U) \right) = \int_0^\infty (\psi(t)-\psi(t_0))(g(t)-f(t))\,\mathrm{d}t\ge0, \]
since the integrand is non-negative.
\end{proof}

\subsection{Metric enlargements}

For a metric $d$ and set $A\subset\mathbb{R}^n$, define as before
\(A_h \coloneqq A_{{d,h}}=\{x:d(x,y)<h\ \text{for some }y\in A\}. \)
\begin{lemma}\label{lem: Minkowski sum formulation of h enlargement}
For
\( d(x,y) \coloneqq 
d_{\varepsilon,\Delta}(x,y)
=
\varepsilon\lceil\|x-y\|/\Delta\rceil
\),
the $h$-enlargement satisfies
\[A+\alpha(h)\mathbb{B}\subseteq A_h\subseteq A+\overline{\alpha(h)\mathbb{B}},\]
where $\alpha(h)\coloneqq\Delta\big(\lceil h/\varepsilon\rceil-1\big)$.
If $A$ is open in the Euclidean topology, then $A_h=A+\alpha(h)\mathbb{B}$.
\end{lemma}
{
Note that if $A$ is not open in the {Euclidean} topology the inclusion $A + \alpha(h) \mathbb{B} \subseteq A_h$ can be strict.  For example and simplicity take $n = \Delta = \varepsilon = 1$.  Let $\|x - y\|= |x-y|$  be the usual absolute value for $x,y \in \mathbb{R}$.  Then, with $A = \{0 \}$ (a closed set in the usual topology, but open with respect to $d$), and $h = \frac 3 2$, 
$$
    A_h = \{0 \}_{\frac 3 2} = [-1,1]
$$
while $\alpha(3/2) = 1$ so that 
$$A+ \alpha(h) \mathbb{B} =\{0 \} +  \mathbb{B} = (-1,1).$$}
\begin{proof}
    For $x \in A + \alpha(h) \mathbb{B}$, by definition there exists $a \in A$ and $y$ with $\|y \| < \alpha(h)$  such that $x = a+y$, hence
    \[
         d(x,a) = \varepsilon \left \lceil \frac{\|x -a\|}{\Delta} \right \rceil \leq \varepsilon \left \lceil \frac{\alpha(h)}{\Delta} \right \rceil.
    \]
    By the definition of $\alpha(h) = \Delta \left( \left\lceil \frac{h}{\varepsilon} \right\rceil - 1 \right)$ the result follows, since
    \begin{align*}
        \varepsilon \left \lceil \frac{\alpha(h)}{\Delta} \right \rceil = \varepsilon \left( \left\lceil \frac{h}{\varepsilon}\right \rceil -1  \right) < \varepsilon \left( \frac{h}{\varepsilon} + 1 -1 \right) = h.
    \end{align*}
    For the second inclusion, $x \in A_h$ gives by definition the existence of $a \in A$ such that
    \[
        d(x,a) = \varepsilon \left \lceil \frac{ \|x-a\|}{\Delta} \right \rceil < h
    \] which yields
    \[
        \frac{ \|x - a\|}{\Delta} \leq \left \lceil \frac{h}{\varepsilon} \right \rceil -1
    \]
    so that setting $y = x -a$, we have $\|y \| \leq \alpha(h)$, and hence $x \in A + \overline{ \alpha(h) \mathbb{B} }$.  
    We now show that $A_h \subseteq A + \alpha(h) \mathbb{B}$ when $A$ is open, so that $A_h = A + \alpha(h) \mathbb{B}$ in this case. For $x \in A_h$, there exists $a \in A$ such that $d(x,a) \leq \alpha(h)$. Moreover, by $A$ open, there exists $\delta > 0$ such that for $t \in (0,\delta)$, $a_t$ defined by
    \[
        a_t \coloneqq (1-t)a + tx 
    \]
    belongs to $A$ and since
    \[
        \| x - a_t \| = (1-t) \|x -a \| < \alpha(h),
    \]
    writing $y \coloneqq x - a_t$, we have shown that $y \in \alpha(h) \mathbb{B}$ and $a_t \in A$.  Thus $x = a_t +y \in A + \alpha(h) \mathbb{B}.$
\end{proof}

\subsection{Proof of Theorem~\ref{Prop:NormRearrangementSmaller}}

\begin{proof}
Let $X$ be such that $M_X$ is $\varepsilon$-DP.
By Theorem~\ref{Thm:DPequivalentEnlarged}, the density $f$ of $X$ satisfies
\[\{\log f>\lambda\}_{d,h}\subseteq\{\log f>\lambda-h\},\]
for all \replaced{$h > 0$ and $\lambda \in \mathbb{R}$ }{$\lambda, h>0$}, where $d(x,y) =  d_{\varepsilon,\Delta}(x,y) = \varepsilon \Big\lceil \tfrac{\|x-y\|}{\Delta} \Big\rceil $.

Let $f^\ast$ be the rearrangement of $f$. Using Proposition~\ref{Prop:PropertiesRearrangements}(ii) and Lemma~\ref{lem: Minkowski sum formulation of h enlargement} with the fact that the rearrangement of a set is an {\it open} ball we can write 
\[{\{\log f^\ast>\lambda\}_{d,h} = \{ f^\ast> e^\lambda \}_{d,h} = (\{ f > e^\lambda\}^\ast)_{d,h} =\{f>e^\lambda\}^\ast+\alpha(h)\mathbb{B}.} \]
By Brunn-Minkowski (see Proposition~\ref{Prop:PropertiesRearrangements}(v)) and Lemma~\ref{lem: Minkowski sum formulation of h enlargement}
\[
|\{f>e^\lambda\}^\ast+\alpha(h)\mathbb{B}|
\le
|\{f>e^\lambda\}+\alpha(h)\mathbb{B}| \leq | \{ f > e^{\lambda} \}_{d,h}|.
\]
By the assumption of the level-set inclusion formulation of Theorem~\ref{Thm:DPequivalentEnlarged} of the assumed log-Lipschitz hypothesis
\(
\{f>e^\lambda\}_{d,h} \subseteq\{f>e^{\lambda-h}\}.
\)
Now since $f$ and $f^\ast$ are equimeasurable we arrive at
\[ {
    |\{ \log f^\ast > \lambda \}_{d,h}| \leq |\{ f > e^{\lambda}\}_{d,h}| \leq |
        \{ f> e^{\lambda - h} \}| = | \{ \log f > \lambda - h \}| = | \{ \log f^\ast > \lambda - h \}|.}
\]
However both sets are open origin symmetric balls, so a volume inequality implies a set-theoretic inclusion.  Thus,
\[\{\log f^\ast>\lambda\}_h\subseteq\{\log f^\ast>\lambda-h\}, \]
and $f^\ast$ satisfies the DP level-set condition, and therefore  Theorem~\ref{Thm:DPequivalentEnlarged} implies that $M_{X^\ast}$ is $\varepsilon$-DP.

It remains to prove that $\|X^\ast\|\prec\|X\|$. To do so, notice that by definition,
$$\Pr(\|X^\ast\|\le r)=\int_{r \mathbb{B}} f^\ast(x)\,\mathrm{d}x,$$
and 
\[\Pr(\|X\|\le r)=\int_{r \mathbb{B}} f(x)\,\mathrm{d}x.\]

Using the layer-cake representation and Proposition~\ref{Prop:PropertiesRearrangements}(ii),
\begin{align*}
\int_{r \mathbb{B}} f^\ast(x)\,\mathrm{d}x
&=
\int_0^\infty
\big|\{f^\ast>t\}\cap r \mathbb{B}\big|\,\mathrm{d}t
\\
&=
\int_0^\infty
\big|\{f>t\}^\ast\cap r \mathbb{B}\big|\,\mathrm{d}t.
\end{align*}
Since $ r \mathbb{B}$ is already a centered $\|\cdot\|$-ball applying Proposition~\ref{Prop:PropertiesRearrangements}(iv) to 
$A=\{f>t\}$ and $B= r \mathbb{B}$ yields
\[
\big|\{f>t\}^\ast\cap r \mathbb{B}\big|
\ge
\big|\{f>t\}\cap r \mathbb{B}\big|
\quad\text{for all }t\ge0.
\]
Integrating over $t$ gives
\[\int_{ r \mathbb{B}} f^\ast(x)\,\mathrm{d}x \ge \int_{ r \mathbb{B}} f(x)\,\mathrm{d}x, \]
and therefore
\[ \Pr(\|X^\ast\|>r)\le \Pr(\|X\|>r),\]
for all $r>0$. This proves $\|X^\ast\|\prec\|X\|$ and completes the proof.
\end{proof}

\section{Proof of Proposition~\ref{Prop:fastrhonorm}}\label{App:PropositionRadial}
Let $f$ be a probability density on $\bR^n$ and let $f^\ast$ be its radially symmetric
decreasing rearrangement with respect to the unit $\|\cdot\|$-ball $\mathbb{B}$.
By Proposition~\ref{Prop:PropertiesRearrangements}(ii), for every $\lambda>0$,
\[
\{f^\ast>\lambda\}=\{f>\lambda\}^\ast.
\]
By Definition~\ref{Def:ConvexRearrangement}, the right-hand side is an (open) centered
$\|\cdot\|$-ball whose radius is determined by volume.
{Thus, for $x,y$ such that $\|x \| = \|y\|$ we have $f^\ast(x) > \lambda$ if and only if $x \in \{ f > \lambda \}^\ast$ if and only if $y \in \{ f > \lambda \}^\ast$ if and only if $f^\ast(y) > \lambda$, and hence $f^\ast(x) = f^\ast(y)$.  Take $u \in \mathbb{R}^n$ such that $\|u\| = 1$, and define $\rho(t) \coloneqq f^\ast(t u)$.  Thus $\rho(\|x\|) = f^\ast( \|x\| u) = f^\ast(x)$ since $\left\| \|x\| u \right\| = \|x\|$, therefore $f^\ast(x)=\rho(\|x\|)$, proving~\eqref{eq:fastrhonorm}.} \added{The decreasingness of $\rho$ follows immediately from the radial decreasingness of $f^*$ as does the openness of $\{ \rho > \lambda \}$ from the openness of $\{ f^* > \lambda \}$}

It remains to relate the DP guarantee of $f^\ast$ to that of $\rho$.
By Appendix~\ref{sec: ep DP is Lipschitz Condition} (Theorem~\ref{Thm:DPequivalentEnlarged}),
the DP constraint for $f^\ast$ is equivalent to $\log f^\ast$ being $1$-Lipschitz
with respect to
\[
d_{\varepsilon,\Delta}(x,y)=\varepsilon\Big\lceil \tfrac{\|x-y\|}{\Delta}\Big\rceil .
\]
Assume first that $\log f^\ast$ is $1$-Lipschitz with respect to $d_{\varepsilon,\Delta}$.
Fix $r,s\ge0$ and choose any $u\in\bR^n$ with $\|u\|=1$. Set $x=ru$ and $y=su$. Then
$\|x\|=r$, $\|y\|=s$, and by homogeneity of the norm, $\|x-y\|=\|(r-s)u\|=|r-s|$, so
\[
d_{\varepsilon,\Delta}(x,y)
=\varepsilon\Big\lceil \tfrac{|r-s|}{\Delta}\Big\rceil
=d_{\varepsilon,\Delta}^{(1)}(r,s).
\]
Using $f^\ast(ru)=\rho(r)$ and $f^\ast(su)=\rho(s)$, the Lipschitz property gives
\[
|\log\rho(r)-\log\rho(s)|
=
|\log f^\ast(x)-\log f^\ast(y)|
\le d_{\varepsilon,\Delta}(x,y)
=
d_{\varepsilon,\Delta}^{(1)}(r,s),
\]
so $\log\rho$ is $1$-Lipschitz with respect to $d_{\varepsilon,\Delta}^{(1)}$.

Conversely, assume that $\log\rho$ is $1$-Lipschitz with respect to
$d_{\varepsilon,\Delta}^{(1)}$. For arbitrary $x,y\in\bR^n$, set $r=\|x\|$ and
$s=\|y\|$. Then
\[|\log f^\ast(x)-\log f^\ast(y)| =|\log\rho(r)-\log\rho(s)|\le \varepsilon\Big\lceil \tfrac{|r-s|}{\Delta}\Big\rceil.\]
By the reverse triangle inequality, $|r-s|=|\|x\|-\|y\||\le \|x-y\|$, hence
\[\Big\lceil \tfrac{|r-s|}{\Delta}\Big\rceil \le \Big\lceil \tfrac{\|x-y\|}{\Delta}\Big\rceil, \]
and therefore
\[|\log f^\ast(x)-\log f^\ast(y)|
\le\varepsilon\Big\lceil \tfrac{\|x-y\|}{\Delta}\Big\rceil
=d_{\varepsilon,\Delta}(x,y). \]
Thus $\log f^\ast$ is $1$-Lipschitz with respect to $d_{\varepsilon,\Delta}$.

\section{Proof of Lemma \ref{lem: X majorizes a Y in D}}
\label{sec: Lemma 1}

The proof of Lemma \ref{lem: X majorizes a Y in D} relies on several auxiliary results. Motivated by Proposition~\ref{Prop:fastrhonorm}, we establish these results by working directly with radial profiles $\rho$. 

\begin{definition}\label{def:rho_y}
Let $\varepsilon,\Delta>0$, let $\rho:[0,\infty)\to[0,\infty)$, and fix $y\ge0$.
Define
\[q_y(r) \coloneqq e^{-\varepsilon \left\lfloor \frac{r-y}{\Delta}\right\rfloor}
\, \rho\!\left(r-\Delta\left\lfloor \tfrac{r-y}{\Delta}\right\rfloor\right).\]
\end{definition}

Equivalently, $q_y$ is uniquely determined by the conditions $q_y(t)=\rho(t)$ for $t\in[y,y+\Delta)$ and $q_y(r+\Delta)=e^{-\varepsilon}q_y(r)$ for all $r\ge0$.

The following proposition shows that $q_y$ will inherit useful properties from $\rho$.

\begin{proposition}\label{prop: rho properties}
Assume that $\rho$ is decreasing, lower semi-continuous, and satisfies
\begin{equation}\label{eq:rho_loglip}
\log\rho(t)-\log\rho(s)\le\varepsilon,
\end{equation}
for $|t-s|\le\Delta$. Then, for $y \in [0,\infty)$, $q_y$ is decreasing, lower semi-continuous, and also satisfies
\eqref{eq:rho_loglip}. Moreover, $\rho(r)\le q_y(r)$ for $r\le y$, $\rho(r)\ge q_y(r)$
for $r\ge y$, and $y \mapsto q_y(r)$ is an increasing function. 
\end{proposition}

\begin{proof}
The construction enforces maximal decay
$q_y(r+\Delta)=e^{-\varepsilon}q_y(r)$ while agreeing with $\rho$ on
$[y,y+\Delta)$, from which monotonicity and the log-Lipschitz bound follow by
induction. The comparison with $\rho$ on either side of $y$ is immediate from the
definition. { For lower semi-continuity, we observe that $\liminf_{r \to t} q_y(r) \geq q_y(t)$ is immediately inherited from the lower semi-continuity of $\rho$ for points $t \notin \{ y + \Delta k\}_{k \in \mathbb{Z}}$.  By the periodicity of $q_y$ it thus suffices to check that $\liminf_{r \to y} q_y(r) \geq q_y(y) = \rho(y).$  This is also immediate since $r < y$ implies $q_y(r) \geq q_y(y)$ from the monotonicity of $q_y$, and for $r \in [y,y+\Delta)$, $q_y(r) = \rho(r)$.}  

To prove the monotonicity of $y \mapsto q_y(r)$, first observe that $y \mapsto\left \lfloor \tfrac{ r - y}{\Delta} \right \rfloor$ is piecewise constant on the intervals $y \in \left( r - (m+1) \Delta , r - m \Delta \right]$ so it suffices to prove $q_{r- m \Delta} (r) \geq q_{r- (m+1) \Delta}(r)$ for integer $m$.  Using $\rho(t + \Delta) \geq e^{-\varepsilon} \rho(t)$, 
\begin{align*}
    q_{r - m \Delta}(r) 
        &=
            e^{- \varepsilon m} \rho \left( r - m\Delta  \right)
                \\
        &=
            e^{-\varepsilon m} \rho(r - (m+1) \Delta + \Delta)
                \\
        &\geq
            e^{-\varepsilon(m+1)} \rho(r - (m+1) \Delta)
                \\
        &=
            q_{r- (m+1)\Delta} (r),
\end{align*}
completes the proof.
\end{proof}

For technical reasons we will need the following simple bound.

\begin{proposition}\label{prop: boundedness of rho functions}
Assume $\rho$ is decreasing. Then for every $y\ge 0$,
\[\|q_y\|_\infty\le e^{\varepsilon \left\lceil \frac{y}{\Delta}\right\rceil}\rho(y)
\le e^{\varepsilon \left\lceil \frac{y}{\Delta}\right\rceil}\rho(0). \]
In particular,
\[
\|q_y\|_\infty \le e^{\varepsilon\left(1+\frac{y}{\Delta}\right)}\rho(0).
\]
\end{proposition}
\begin{proof}
By Proposition~\ref{prop: rho properties}, $q_y$ is decreasing on $[0,\infty)$,
hence $\|q_y\|_\infty=q_y(0)$. Using the definition of $q_y$ at $r=0$,
\[
q_y(0)
=
e^{-\varepsilon \left\lfloor \frac{-y}{\Delta}\right\rfloor}
\,
\rho\!\left(-\Delta\left\lfloor \tfrac{-y}{\Delta}\right\rfloor\right).
\]
Since $\lfloor -a\rfloor = -\lceil a\rceil$, we get
\[
q_y(0)
=
e^{\varepsilon \left\lceil \frac{y}{\Delta}\right\rceil}
\,
\rho\!\Big(\Delta\Big\lceil \tfrac{y}{\Delta}\Big\rceil\Big).
\]
Moreover, $\Delta\lceil y/\Delta\rceil \in [y,y+\Delta)$, so by monotonicity of $\rho$,
\[
\rho\!\Big(\Delta\Big\lceil \tfrac{y}{\Delta}\Big\rceil\Big)\le \rho(y)\le \rho(0),
\]
which yields the claimed bound. Finally, $\lceil y/\Delta\rceil\le y/\Delta+1$ implies
$\exp(\varepsilon\lceil y/\Delta\rceil)\le \exp(\varepsilon(1+y/\Delta))$.
\end{proof}

\begin{lemma}\label{lem: maximally decreasing radial density}
Let $\rho:[0,\infty)\to[0,\infty)$ be lower semi-continuous and decreasing, satisfy
\begin{equation}\label{eq: log-lip is sat C}
\log\rho(t)-\log\rho(s)\le \varepsilon,
\end{equation}
for $|t-s|\le \Delta$, and assume that $r\mapsto r^{n-1}\rho(r)$ is integrable for an integer $n$. Then there exists $y\ge0$ such that
\[
\int_0^\infty r^{n-1}\rho(r)\,\mathrm{d}r
=
\int_0^\infty r^{n-1}q_y(r)\,\mathrm{d}r,
\]
for $q_y$ as in Definition \ref{def:rho_y}.
\end{lemma}
\begin{proof}

Define $\psi(y)\coloneqq\int_0^\infty r^{n-1}q_y(r)\,\mathrm{d}r$ and recall 
\[q_0(r) =e^{-\varepsilon \left\lfloor \frac{r}{\Delta} \right\rfloor}\rho\!\left(r - \Delta \left\lfloor \tfrac{r}{\Delta} \right\rfloor\right).\]
By construction, $q_0$ enforces the maximal decay
$q_0(r+\Delta)=e^{-\varepsilon}q_0(r)$ for all $r\ge 0$ while agreeing with $\rho$ on $[0,\Delta)$. We next show that the log-Lipschitz condition~\eqref{eq: log-lip is sat C} implies $q_0(r)\le\rho(r)$ for all $r\ge0$. To see this, note that \eqref{eq: log-lip is sat C} directly implies  $\rho(t+\Delta)\ge e^{-\varepsilon}\rho(t)$ for all $t\ge 0$, and by iterating it yields $\rho(u+k\Delta)\ge e^{-k\varepsilon}\rho(u)$ for every $u\in[0,\Delta)$ and $k\ge 0$. Writing $r=k\Delta+u$, this shows
\[\rho(r)\ge e^{-k\varepsilon}\rho(u)=q_0(r), \]
and hence $q_0\le\rho$ pointwise on $[0,\infty)$. This means $\psi(0)\le\int_0^\infty r^{n-1}\rho(r)\,\mathrm{d}r$.

If $\rho$ itself already satisfies the exact decay identity
$\rho(r+\Delta)=e^{-\varepsilon}\rho(r)$ for all $r\ge 0$, then $q_y=\rho$ for every
$y\ge 0$, and the desired equality holds trivially.
Otherwise, there exists a point $t_0 > 0$ at which the decay constraint is not tight,
that is,
\[\rho(t_0)\;<\;e^{\varepsilon}\rho(t_0+\Delta). \]
By lower semi-continuity of $\rho$, this strict inequality persists on a neighborhood of
$t_0$: there exist $\eta>0$ and $\delta>0$ such that
\[
t^{n-1}\rho(t)\;<\;e^{\varepsilon} t^{n-1}\rho(t+\Delta)-\delta
\qquad\text{for all }t\in[t_0,t_0+\eta].
\]
By Proposition~\ref{prop: rho properties} for $y \geq t_0 + \eta + \Delta$ and $t \in [t_0, t_0 + \eta]$,
\[
    t^{n-1} q_y(t) = e^{\varepsilon} t^{n-1} q( t + \Delta) \geq e^{\varepsilon} t^{n-1} \rho(t + \Delta) \geq t^{n-1} \rho(t) + \delta.
\]
Applying this inequality and again using Proposition~\ref{prop: rho properties} that $q_y \geq p$ on $[0, y]$,
\begin{align*}
    \psi(y) - \int_0^\infty r^{n-1} \rho(r)dr
        &=
            \int_0^y r^{n-1} (q_y(r) - \rho(r)) dr + \int_y^\infty r^{n-1} (q_y(r)- \rho(r) dr
                \\
        &\geq
            \int_{t_0}^{t_0 + \eta} r^{n-1}(q_y(r) - \rho(r)) dr - \int_y^\infty r^{n-1} \rho(r) dr
                \\
        &\geq
            \eta \delta - \int_y^\infty r^{n-1} \rho(r) dr.
\end{align*}
Taking $y \to \infty$ and using the integrability of $r^{n-1} \rho(r)$, 
\[
    \limsup_{y \to \infty} \psi(y) \geq \int_0^\infty r^{n-1} \rho(r) dr  + \eta \delta.
\]


To conclude we need only show that $\psi$ is continuous. Let $y_k\to y$ with $M \coloneqq \sup_{k} y_k$. By Proposition~\ref{prop: rho properties} for $r \leq M$, $$r^{n-1} q_{y_k}(r) \leq M^{n-1} q_M(r) \leq M^{n-1} e^{\varepsilon \left \lceil \tfrac{M}{\Delta} \right \rceil} \rho(0) < \infty$$ 
and on $[M, \infty)$ $r^{n-1} q_{y_k} (r) \leq r^{n-1} \rho(r)$. As it is easily verified that $q_{y_k} \to q_y$ almost everywhere, dominated convergence therefore implies $\psi(y_k)\to\psi(y)$.

Thus the intermediate value
theorem guarantees the existence of some $y\ge 0$ such that
\[
\psi(y)=\int_0^\infty r^{n-1}\rho(r)\,\mathrm{d}r,
\]
which completes the proof.
\end{proof}

We can now present the proof of Lemma~\ref{lem: X majorizes a Y in D}. 


\begin{proof}[Proof of Lemma~\ref{lem: X majorizes a Y in D}]
Let $X\sim\mu\in\mathcal{M}_n(\Delta,\varepsilon)$. By
Theorem~\ref{Prop:NormRearrangementSmaller}, its rearrangement $X^\ast$ satisfies
$\|X^\ast\|\prec\|X\|$. Writing $f_{X^\ast}(x)=\rho(\|x\|)$, Proposition~\ref{Prop:fastrhonorm}
ensures that $\rho$ is decreasing and satisfies~\eqref{eq: log-lip is sat C}.
Choose $y\ge0$ according to Lemma~\ref{lem: maximally decreasing radial density} such that 
 \[
        \int_0^\infty r^{n-1}\rho(r) \mathrm{d}r = \int_0^\infty r^{n-1}  q_y(r)\mathrm{d}r. 
    \]
This guarantees that $ q_y(\|x\|)$ is a probability density. Thus, we define $f_Y(x)\coloneqq q_y(\|x\|)$. Moreover, Proposition~\ref{prop: rho properties} shows that
$ q_y(r)\ge\rho(r)$ for $r\le y$ and $ q_y(r)\le\rho(r)$ for $r\ge y$, so the radial
densities of $\|Y\|$ and $\|X^\ast\|$ cross exactly once. Lemma~\ref{Lemma:DensityStochasticDomination}
therefore yields $\|Y\|\prec\|X^\ast\|$. Since $\|X^\ast\|\prec\|X\|$, transitivity of
stochastic domination completes the proof.
\end{proof}

\section{Proof of Theorem~\ref{thm: extreme points of the space D}}  \label{sec: extreme points}
In this section, we establish, using functional-analytic arguments, that for any norm-monotone cost function $\Phi$, an optimal mechanism in $\mathcal{M}_n(\Delta,\varepsilon)$ may be chosen to be a
staircase mechanism. The proof relies on an application of the Krein--Milman theorem.

\begin{theorem}[Krein--Milman]\label{thm: Krein Milman}
A compact convex subset of a locally convex Hausdorff topological vector space is equal to the closed
convex hull of its extreme points.
\end{theorem}

A standard reference for Krein--Milman and related results is \cite{phelps2001lectures}.
The purpose of Theorem~\ref{thm: extreme points of the space D} is to verify that the hypotheses of
Theorem~\ref{thm: Krein Milman} apply to the class $\mathcal{D}$ introduced in
Definition~\ref{def: The space D}. Here, the space of finite signed measures with the weak topology will play the role of the Hausdorff locally convex topological vector space.
Throughout this section, compactness is understood with respect to the weak topology on finite signed
measures, i.e., $\mu_k \to \mu$ if and only if
\[\int f\,\mathrm{d}\mu_k \;\to\; \int f\,\mathrm{d}\mu, \]
for all bounded continuous functions $f$. 
The proof of Theorem~\ref{thm: extreme points of the space D} proceeds in two steps:
\begin{enumerate}
\item show that $\mathcal{D}$ is convex and weakly compact;
\item characterize the extreme points of $\mathcal{D}$.
\end{enumerate}

We begin with a useful comparison between elements of $\mathcal{D}$ and the Laplace distribution.
\begin{lemma}\label{lem: control by Laplace}
Let $\mu\in\mathcal{D}$ have density $f(x)=\rho(\|x\|)$. Then
\[e^{-2\varepsilon}\ell(t)\le \rho(t) \le e^{2\varepsilon}\ell(t), \qquad t\ge0,\]
where
\begin{equation}
    \ell(t)=\frac{\varepsilon^n}{|\mathbb{B}|\,n!\,\Delta^n}\, e^{-\varepsilon t / \Delta}.
\end{equation}
In particular, for $\mu_i \in \mathcal{D}$ with densities $f_i$, we have
\[
    f_0 \leq e^{4 \varepsilon} f_1.
\]
\end{lemma}
Note that the function  $\ell(t)$  corresponds to the radially symmetric density
\[f_\ell (x)=\frac{\varepsilon^n}{|\mathbb{B}|\,n!\,\Delta^n}\exp\!\left(-\frac{\varepsilon}{\Delta}\|x\|\right), \]
that can be viewed as the radial analogue of the classical product Laplace mechanism for general norm $\|\cdot\|$. Since the logarithm of $f$ is $1$--Lipschitz with respect to the metric $d_{\varepsilon,\Delta}$, it defines an additive mechanism in
$\mathcal{M}_n(\Delta,\varepsilon)$. When $\|\cdot\|=\|\cdot\|_1$, the product Laplace mechanism has density proportional to $\exp(-(\varepsilon/\Delta)\|x\|_1)$ and is itself radial in the $\ell_1$ geometry; for general norms, the density above provides the canonical radial extension.

\begin{proof}
Since $\rho$ is decreasing and satisfies $\rho(\Delta)=e^{-\varepsilon}\rho(0)$, the log-Lipschitz
condition implies
\begin{equation}\label{eq: pointwise bound needed}
    e^{-\varepsilon}\rho(0)e^{-\varepsilon t/\Delta}\le \rho(t)
\le e^{\varepsilon}\rho(0)e^{-\varepsilon t/\Delta},
\end{equation}
for $t\in[0,\Delta]$.
By the maximal decay relation $\rho(t+\Delta)=e^{-\varepsilon}\rho(t)$, the same inequality holds for
all $t\ge0$.
Multiplying by $|\mathbb{B}|\,n\,t^{n-1}$ and integrating over $[0,\infty)$ yields
\[e^{-\varepsilon}\rho(0)\,|\mathbb{B}|\,n!\Big(\frac{\Delta}{\varepsilon}\Big)^n \le1\le e^{\varepsilon}\rho(0)\,|\mathbb{B}|\,n!\big(\frac{\Delta}{\varepsilon}\Big)^n,\]
from which 
\[ \frac{e^{-\varepsilon}\varepsilon^n}{|\mathbb{B}|\,n!\,\Delta^n}
\le\rho(0)\le \frac{e^{\varepsilon}\varepsilon^n}{|\mathbb{B}|\,n!\,\Delta^n}.\]
Substituting this bound back into the pointwise estimate \eqref{eq: pointwise bound needed} completes the proof.  To confirm the particular case, let $\rho_i$ denote the radial function of $f_i$, and applying the comparison against $\ell$ to both $\rho_i$,
\[
    \rho_0(t) \leq e^{2\varepsilon} \ell(t) \leq  e^{4 \varepsilon} \rho_1(t).
\]
Thus $f_0(x) = \rho_0(\|x\|) \leq e^{4 \varepsilon} \rho_1(\|x\|) = e^{4 \varepsilon} f_1(x).$
\end{proof}

{
\begin{corollary} \label{cor: continuity of F}
    Given a norm-monotone cost function $\Phi$, define $F: \mathcal{D} \to [0,\infty]$ by
    \[
        F(\mu) = \int \Phi \ \mathrm{d}\mu.
    \]
    If there exists $\mu_1 \in \mathcal{D}$ such that $F(\mu_1) <\infty$, then $F(\mu) < \infty$ for all $\mu \in \mathcal{D}$ and, moreover, $F$ is continuous on $\mathcal{D}$ with respect to the weak topology.
\end{corollary}

\begin{proof}
    Given $\mu_0 \in \mathcal{D}$ with density function $f_0$ and letting $f_1$ be the density function of $\mu_1$, by Lemma \ref{lem: control by Laplace} we have { that $f_0 \leq e^{4 \varepsilon} f_1$ pointwise and hence}
    \[
        F(\mu_0) = \int \Phi f_0 \leq e^{4 \varepsilon} \int \Phi f_1 = e^{4 \varepsilon} F(\mu_1) < \infty.
    \]
Let $\mu_k \to \mu$ in the weak topology, and let $\rho_k$ and $\rho$ denote the radial functions corresponding to the density functions $f_k$ and $f$.  It is easy to see that $\rho_k(t) \to \rho(t)$ if $\rho$ is continuous at $t$.  Indeed, $\limsup \rho_k(t) \leq \rho(t)$, else there exists an $\eta >0$ and a subsequence $\rho_{k_j}(t)$ such that $\rho_{k_j}(t) > \rho(t) +  2 \eta$. By continuity of $\rho$ at $t$, there exists a $\delta > 0$ such that $\rho (t - \delta) < \rho(t) + \eta$, hence for $s \in [t - \delta, t]$,
    \[
        \rho_{k_j}(s) \geq \rho_{k_j}(t) \geq \rho(t) + 2 \eta \geq \rho(t-\delta) + \eta \geq \rho(s) + \eta,
    \]
    from which it is easy to see that weak convergence would be contradicted by considering the continuity set for $\mu$, $t \mathbb{B} \setminus (t- \delta) \mathbb{B}$.  The argument for $\liminf \rho_k(t) \geq \rho(t)$ is similar\footnote{In fact $\liminf \rho_k(t) \geq \rho(t)$ holds at all points $t$ by the lower semi-continuity of $\rho$.}.  Thus, since $\rho$ is a decreasing function, it is continuous at all but countably many points, and it thus follows that $f_k \to f$ almost everywhere. By Lemma \ref{lem: control by Laplace}, $\Phi f_k \leq e^{4 \varepsilon} \Phi f$, and since the latter is integrable one can apply dominated convergence,
    \[
        \lim_{k \to \infty} F(\mu_k) = \lim_{k\to \infty} \int \Phi f_k = \int \Phi \lim_{k \to \infty} f_k = \int \Phi f = F(\mu).
    \]
\end{proof}

}

\begin{theorem}[Convexity and compactness of $\mathcal{D}$]\label{lem: convex and compact}
The set $\mathcal{D}$ is convex and compact in the weak topology.
\end{theorem}
\begin{proof}
We prove convexity directly and establish compactness by showing that
$\mathcal{D}$ is tight and weakly closed, allowing us to invoke Prokhorov’s theorem.

Convexity is immediate: if $\mu_i\in\mathcal{D}$ have densities
$f_i(x)=\rho_i(\|x\|)$, then any convex combination $f=(1-\lambda)f_1+\lambda f_2$ has radial function $\rho=(1-\lambda)\rho_1+\lambda\rho_2$ and satisfies the defining properties of $\mathcal{D}$.

To prove compactness, we first show that $\mathcal{D}$ is tight.
By Lemma~\ref{lem: control by Laplace}, there exists a constant
$C<\infty$, depending only on $(\varepsilon,\Delta,n)$, such that for all
$\mu\in\mathcal{D}$,
\begin{align*}
\mu(\|x\|>t)
&=\int_t^\infty \rho(r)\, n r^{n-1}|\mathbb{B}|\,\mathrm{d}r \\
&\le C\int_t^\infty r^{n-1}e^{-\varepsilon r/\Delta}\,\mathrm{d}r .
\end{align*}
This uniform tail bound implies tightness. By Prokhorov’s theorem
(see, e.g., \cite{billingsley2013convergence}), $\mathcal{D}$ is therefore
precompact in the weak topology.

It remains to show that $\mathcal{D}$ is weakly closed.
Let $\mu_k\in\mathcal{D}$ converge weakly to $\mu$, written $\mu_k  \xrightarrow{w} \mu$.
Lemma~\ref{lem: control by Laplace} implies that the densities of $\mu_k$
are uniformly bounded, and hence $\mu$ is absolutely continuous with a
bounded density $f$. { Indeed, by the uniform boundedness there exists a constant $C > 0$
such that 
\[
    \mu_k(E) \leq C |E|,
\]
for any Borel set $E$.  Now fix $E$ and take $A$ to be an open set such that $E \subseteq A$,
By weak convergence $\liminf \mu_k(A) \geq \mu(A)$, hence
\begin{align} \label{eq: weak convergence preserves infinity norms}
    \mu(E) \leq \mu(A) \leq \liminf_{k \to \infty} \mu_k(A) \leq C |A|. 
\end{align}
Since the Lebesgue measure is outer regular taking the infimum over all such $A$, we have $\mu(E) \leq C |E|$ and absolute continuity of $\mu$ follows.}

For $X_k\sim\mu_k$, since $X_k  \xrightarrow{w} X$ and $x \mapsto \|x\|$ is continuous, $\|X_k\| \xrightarrow{w} \|X\|$.  By assumption, $\|X_k\|$ has a density function on $(0,\infty)$
\[
    g_k(t) = c \ t^{n-1} \rho_k(t)
\]
with $c = n | \mathbb{B}|$.  Since the $\rho_k(t)$ are uniformly sub-exponential, $g_k$ are uniformly bounded.  Thus by the same argument detailed in \eqref{eq: weak convergence preserves infinity norms}, $\|X\|$ has a density function $g$.  By Lebesgue differentiation theorem we can  write
\[
    g(t) \coloneqq \limsup_{s \downarrow 0} \frac{\int_{t}^{t+s} g(w) dw}{s},
\]
for $t \in (0,\infty)$.
Define $f(x) = \rho(\|x\|)$ where
\[
    \rho(t) \coloneqq \frac{g(t)}{c \ t^{n-1}}.
\]
We claim that $\rho(t)$ is decreasing.  Consider $t_0 < t$ and $\eta >0$,
\begin{align*}
    \rho(t)
        &=
           \limsup_{s \downarrow 0} \frac{\int_{t}^{t+s} g(w) dw}{s \ c \ t^{n-1}} 
                \\
        &=
            \limsup_{s \downarrow 0} \frac{ \lim_{k \to \infty} \int_{t}^{t+s} g_k(w) dw}{s \ c \ t^{n-1}} 
                \\
        &=
            \limsup_{s \downarrow 0} \frac{ \lim_{k \to \infty} \int_{t}^{t+s} c \ w^{n-1} \rho_k(w) dw}{s \ c \ t^{n-1}} 
                \\
        &\leq
            \limsup_{s \downarrow 0} \frac{ (t+\eta)^{n-1} \lim_{k \to \infty} \int_{t}^{t+s}   \rho_k(w) dw}{s  \ t^{n-1}} 
                \\
        &\leq
            \left( \frac{t + \eta}{t} \right)^{n-1}\limsup_{s \downarrow 0} \frac{  \lim_{k \to \infty} \int_{t_0}^{t_0+s}   \rho_k(w) dw}{s} 
                \\
        &=
            \left( \frac{t + \eta}{t} \right)^{n-1}\limsup_{s \downarrow 0} \frac{  \lim_{k \to \infty} \int_{t_0}^{t_0+s}   \frac{g_k(w)}{c w^{n-1}} dw}{s}
                \\
        &\leq
            \left( \frac{t + \eta}{t} \right)^{n-1}\limsup_{s \downarrow 0} \frac{  \lim_{k \to \infty} \int_{t_0}^{t_0+s}   g_k(w) dw}{s \ c \ t_0^{n-1}}
                \\
        &=
            \left( \frac{t + \eta}{t} \right)^{n-1}\limsup_{s \downarrow 0} \frac{   \int_{t_0}^{t_0+s}   g(w) dw}{s \ c \ t_0^{n-1}}   \\  
        &=
            \left( \frac{t + \eta}{t} \right)^{n-1} \frac{ g(t_0)}{ c \ t_0^{n-1}}
                \\
        &=
            \left( \frac{t + \eta}{t} \right)^{n-1}  \rho(t_0).
\end{align*}
Line by line, we use, the Lebesgue differentiation theorem representation of $g$, the weak convergence $\|X_k\| \xrightarrow{w} \|X\|$, the expression of $g_k$ in terms of the radial function $\rho_k$ associated to the density of $X_k$, that $r \mapsto r^{n-1}$ is strictly increasing (valid for $s<\eta$), that $\rho_k$ is decreasing, the expression of $g_k$ in terms of $\rho_k$ again, that $r \mapsto r^{n-1}$ is increasing again, the Lebesgue differentiation representation of $g$, and the definition of $\rho$.
Taking $\eta \to 0$ completes the proof.

Now we show that $\rho$ is lower semi-continuous.  Let us first observe that the property that $\rho(t) = \limsup_{s \downarrow 0} \frac{\int_t^{t+s} \rho(r) dr}{s}$ is inherited from $g$.  Indeed, for $\eta >0,$
\begin{align*}
    \limsup_{s \downarrow 0} \frac{\int_t^{t+s} \rho(r) dr}{s} \geq \limsup_{s \downarrow 0}  \frac{\int_t^{t+s} g(r) dr}{c (t + \eta)^{n-1} s} =  \frac{g(t)}{c(t +\eta)^{n-1}}. 
\end{align*}
Taking $\eta \to 0$ shows the limsup is bounded below by $\rho(t)$.  The converse inequality is immediate from the monotonicity of $\rho$ proved above. Now we show that $\{ \rho > \lambda \}$ is an open set.  \replaced{For $t$ such that $\rho(t) > \lambda$, there must exist $\delta > 0$ such that $\rho(t + \delta) > \lambda$, else $\rho(t) = \limsup_{s \downarrow 0} \frac{\int_t^{t+s} \rho(w) dw}{s} \leq \lambda$ and contradiction.  By monotonicity $(0,t + \delta) \subseteq \{ \rho > \lambda \}$ and the openness follows.}{For $\rho(t) > \lambda$, there exists $s_0 >0$ such that} \deleted{ 
\(
   \lambda < \frac{\int_t^{t + s_0} \rho(w) dw}{s_0} 
\)
and hence there exists $s' \in (0,s_0)$ such that
\(
    \rho(t + s') > \lambda.
\)
For $r \leq t + s'$,
\(
    \rho(r) = \limsup_{s \downarrow 0} \frac{\int_{r}^{r + s} \rho(w) dw }{s} \geq \frac{\int_r^{r + s'} \rho(w) dw}{s'} \geq \rho(r + s') > \lambda.
\)}
Thus $\rho$ is lower semi-continuous and decreasing.  
Norm symmetry implies that $\|X_k\|$ and
$X_k/\|X_k\|$ are independent, with $X_k/\|X_k\|$ distributed according
to $\sigma$, the uniform probability measure on the unit sphere
$\{x\in\mathbb{R}^n:\|x\|=1\}$ (see Appendix~\ref{sec: Norm Symmetric Coordinates}).
Weak convergence then implies that the same independence and uniformity
hold for $X\sim\mu$, and therefore
\[f(x)=\rho(\|x\|),\]
for some radial function $\rho$. In other words, the weak limit of
radially symmetric measures is radially symmetric as well. See Lemma~\ref{lem: norm symmetry is indpendence and uniform} for details.

Finally, we verify the maximal decay condition. Let $\nu_k$ and $\nu$
denote the laws of $\|X_k\|$ and $\|X\|$, respectively. Since
$\mu_k \xrightarrow{w}  \mu$ and $x\mapsto \|x\|$ is continuous, we have
$\nu_k \xrightarrow{w}  \nu$. Moreover,
\[
    d\nu_k(r)=c r^{n-1}\rho_k(r)\,dr,
    \qquad
    d\nu(r)=c r^{n-1}\rho(r)\,dr,
    \qquad c=n|\mathbb B|.
\]

We first observe that, for every $0<a<b<\infty$,
\begin{align} \label{eq: passing weak convergence}
    \int_a^b \rho_k(r)\,dr
    =
    \int_{(0,\infty)}
    \frac{\mathbf 1_{[a,b]}(r)}{c r^{n-1}}\,d\nu_k(r)
    \longrightarrow
    \int_{(0,\infty)}
    \frac{\mathbf 1_{[a,b]}(r)}{c r^{n-1}}\,d\nu(r)
    =
    \int_a^b \rho(r)\,dr, 
\end{align}
by weak convergence as $\nu$ is absolute continuous and
\[
    r\longmapsto
    \frac{\mathbf 1_{[a,b]}(r)}{c r^{n-1}}
\]
is bounded and is continuous outside of the points $a$ and $b$.

Fix $t>0$ and $\eta>0$. The defining relation
$\rho_k(r+\Delta)=e^{-\varepsilon}\rho_k(r)$ gives
\[
    \int_{t+\Delta}^{t+\Delta+\eta}\rho_k(r)\,dr
    =
    e^{-\varepsilon}
    \int_t^{t+\eta}\rho_k(r)\,dr.
\]
Passing to the limit $k\to\infty$ by \eqref{eq: passing weak convergence} gives
\[
    \int_{t+\Delta}^{t+\Delta+\eta}\rho(r)\,dr
    =
    e^{-\varepsilon}
    \int_t^{t+\eta}\rho(r)\,dr.
\]
Since $\rho$ is decreasing and lower semicontinuous, it is
right-continuous. Dividing by $\eta$ and letting $\eta\downarrow0$
therefore yields
\[
    \rho(t+\Delta)=e^{-\varepsilon}\rho(t),
    \qquad t>0.
\]
Finally, letting $t\downarrow0$ and using right-continuity once more
gives
\[
    \rho(\Delta)=e^{-\varepsilon}\rho(0).
\]
Hence
\[
    \rho(t+\Delta)=e^{-\varepsilon}\rho(t)
\]
for every $t\geq0$. We conclude that $\mu\in\mathcal D$, so
$\mathcal D$ is weakly closed. Combined with tightness, this proves
that $\mathcal D$ is compact in the weak topology.

\end{proof}

\begin{theorem}[Staircase distributions are the extreme points of $\mathcal{D}$]
A measure $\mu\in\mathcal{D}$ is an extreme point of $\mathcal{D}$ if and only if
the lower semi-continuous radial function $\rho$ of its density $f(x)=\rho(\|x\|)$
takes exactly two values on $[0,\Delta]$.
\end{theorem}
Observe that staircase distributions (see Definition~\ref{def:Staircase_mech}) are elements of $\mathcal{D}$, and are characterized by, after taking their lower-semi continuous version if necessary, having a radial function $\rho$ that takes exactly two values on $[0, \Delta]$.  
\begin{proof}
Recall that $\mu\in\mathcal{D}$ means that $\mu$ has a (lower semi-continuous) radial
density $f(x)=\rho(\|x\|)$ where $\rho:[0,\infty)\to[0,\infty)$ is decreasing,
$r\mapsto r^{n-1}\rho(r)$ is integrable, and $\rho$ satisfies the maximal decay rule
\begin{equation}\label{eq:max_decay_extreme}
\rho(t+\Delta)=e^{-\varepsilon}\rho(t),
\end{equation}
for $t\ge 0$. We first show that any $\mu\in\mathcal{D}$ whose radial function $\rho$ takes at least three distinct values on $[0,\Delta]$ is not extreme. The argument is carried out at the level of radial functions.\\

\noindent\emph{($\Rightarrow$) If $\rho$ takes at least three values on $[0,\Delta]$, then $\mu$ is not extreme:}
Assume that $\rho$ takes at least three distinct values on $[0,\Delta]$. Since $\rho$ is decreasing, there exists $s \in (0,\Delta)$ such that
\[
\rho(0)>\rho(s)>\rho(\Delta) = e^{-\varepsilon} \rho(0).
\]
For this point $s$ define the connected\footnote{$\mathcal{S}$ is indeed connected as the finite intersection of halfspaces who clearly have non-empty interior and then removing $0$.} set
\begin{align} \label{eq: equations from S}
\mathcal{S} \;=\; \left\{
  \delta \in \mathbb{R}^2 - \{0\} \;:\;
  \begin{aligned}
    &0 \le \delta_1 \le \min\!\left\{\, \rho(0)-\rho(s),\; e^{\varepsilon}\rho(s)-\rho(0) \right\},\\[2pt]
    &0 \le \delta_2 \le \min\!\left\{\, \rho(0)-\rho(s)-\delta_1,\; \rho(s)- e^{-\varepsilon}(\rho(0)+\delta_1) \right\}.
  \end{aligned}
\right\}
\end{align}
Thus for $\delta \in \mathcal{S}$,  we have
\[\rho(0)-\delta_1 \geq\rho(s)+\delta_2,
\quad ~~\text{and}\quad ~~
\rho(s)\ge e^{-\varepsilon}(\rho(0)+\delta_1)+\delta_2.
\]
The rough idea will be to use the point $s$  to define two segments of interest $[0,s]$ and $[s, \Delta]$ and to increase $\rho_1$ by $\delta_1$ on $[0,s]$ and decrease by $\delta_2$ on $(s, \Delta]$ while doing the opposite for $\rho_2$ and then use the flexibility in $\delta$ to obtain $\rho_i$ that are the radial functions of densities.  However, to ensure that $\rho_i$ belong to $\mathcal{D}$ a more delicate construction is necessary. 
Define for each $\delta \in \mathcal{S}$ the function $\rho_1$ on $[0,\Delta]$ by
\[
\rho_1(t)\coloneqq
\begin{cases}
\rho(t)+\min\{ \delta_1,\ \rho(t) -\rho(s)-\delta_2\},
& t\in[0,s],\\[0.4em]
\rho(t) - \min \{\delta_2,\ \rho(t) -  e^{-\varepsilon} ( \rho(0) + \delta_1) \},
& t\in[s,\Delta].
\end{cases}
\]
and
\[
\rho_2(t)\coloneqq
\begin{cases}
\rho(t) - \min\{\delta_1,\ \rho(t) - \rho(s) -\delta_2\},
& t\in[0,s],\\[0.4em]
\rho(t) + \min\{\delta_2,\ \rho(t) - e^{-\varepsilon} (\rho(0) +\delta_1) \},
& t\in[s,\Delta].
\end{cases}
\]
Then extend the $\rho_i$ to $[0,\infty)$ inductively via $\rho_i(x + \Delta) = e^{-\varepsilon} \rho_i(x)$.
If the $\rho_i$ are well defined, we immediately have $\rho=\tfrac12(\rho_1+\rho_2)$ on $[0,\Delta]$.   Let us first check that the $\rho_i$ are well defined at $s$ and then verify that the $\rho_i$ are well defined at $\Delta$. By direct computation,
\begin{align*}
    \min \{ \delta_1, \rho(s) - \rho(s) - \delta_2 \} = - \delta_2,
\end{align*}
while \eqref{eq: equations from S} gives
\begin{align*}
    - \min \{ \delta_2 , \rho(s) - e^{-\varepsilon}(\rho(0) + \delta_1 ) \} =  - \delta_2
\end{align*}
so the two equations agree at $s$.  The same argument shows that $\rho_2$ is well defined at $s$ as well.  Observe that the definition of $\mathcal{S}$, particularly $\delta_2 \leq \rho(0) - \rho(s) - \delta_1$ ensures that $\rho_1(0) = \rho(0) + \delta_1$ and $\rho_2(0) = \rho(0)- \delta_1$.  Now let us check that the $\rho_i$ are decreasing.  For $\rho_1$ on $[0,s]$ and $\rho_2$ on $[s,\Delta]$ this is direct since the min operation and sums preserve lower semi-continuity and monotonicity.  For $\rho_1$ on $[s, \Delta]$, the representation 
\[
    \rho(t) - \min \{\delta_2,\ \rho(t) -  e^{-\varepsilon} ( \rho(0) + \delta_1) \} = \max \{ \rho(t) - \delta_2,  e^{-\varepsilon} ( \rho(0) + \delta_1) \}
\]
and that the max operation preserves monotonicity and lower semi-continuity delivers $\rho_1$ decreasing and lower semi-continuous on $[s, \Delta]$, and since the two pieces of the definition of $\rho_1$ agree at $s$, $\rho_1$ is decreasing and lower semi-continuous. Similarly
the representation
\[
    \rho(t) - \min\{\delta_1,\ \rho(t) - \rho(s) -\delta_2\} = \max \{ \rho(t) - \delta_1, \rho(s) + \delta_2 \}
\]
of $\rho_2$ for $t \in [0, s]$ gives that $\rho_2$ is decreasing and lower semi-continuous on the interval and the agreement of the piecewise definitions of $\rho_2$ at $s$ gives that $\rho_2$ is decreasing and lower-semi continuous on $[0, \Delta]$.  The monotonicity and lower semi-continuity of the $\rho_i$ immediately extend to the intervals $[k \Delta, (k+1) \Delta]$ via the inductive definition once we know that the $\rho_i$ are well defined at the end points.  For this it suffices to check $t= \Delta$.
\begin{align*}
    \rho_1(\Delta) 
        &= 
            \rho(\Delta) - \min \{ \delta_2, \rho(\Delta) - \rho(\Delta) - e^{-\varepsilon} \delta_1 \}
                \\
        &=
            e^{-\varepsilon}\rho(0) + e^{-\varepsilon} \delta_1
                \\
        &=
            e^{-\varepsilon}( \rho(0) + \delta_1)
                \\
        &=
            e^{-\varepsilon} \rho_1(0)
\end{align*}
Similarly $\rho_2(\Delta) = e^{-\varepsilon} \rho_2(0)$.

Each $\rho_i$ is nonnegative, decreasing, satisfies the maximal decay condition, and hence $\rho_i$ defines an element of $\mathcal{D}$ as soon as 
\[\int_{\mathbb{R}^n} \rho_i(\|x\|) \mathrm{d}x = n |\mathbb{B}| \int_0^\infty \rho_i(r)\,  r^{n-1}\mathrm{d}r = 1 .
\]
Now define $\Theta: \mathcal{S} \to \mathbb{R}$ by
\[ \Theta(\delta)\coloneqq
\int_0^\infty r^{n-1}\bigl(\rho_1(r)-\rho_2(r)\bigr)\,\mathrm{d}r. \]
$\Theta$ is continuous in $\delta$ as the definition of the $\rho_i$ is pointwise continuous in $\delta$, and Lebesgue dominated convergence can be invoked via the fact that $\rho_i \leq 2 \rho$ on $[0,\Delta)$ and hence by the exponential decay condition thereafter $\rho_i \leq 2 \rho$ on $[0,\infty)$. Observe that if $\delta_1=0$ and
$\delta_2>0$, then $\rho_2\ge\rho$ with strict inequality on a neighborhood of
$s$, implying $\Theta(0,\delta_2)<0$. Conversely, if $\delta_2=0$ and
$\delta_1>0$, then $\rho_1\ge\rho$ with strict inequality near $0$, implying
$\Theta(\delta_1,0)>0$. $\Theta(\mathcal{S})$ is the continuous image of a connected set, and hence is connected itself.  Since $\Theta(\mathcal{S})$ contains negative and positive values, it contains $0$ and hence there exists a choice of $\delta$ such that $\Theta(\delta) = 0$.  Thus
\[
    \int_{\mathbb{R}^n} \rho_1(\|x\|) \mathrm{d}x =  \int_{\mathbb{R}^n} \rho_2(\|x\|) \mathrm{d}x = \int_{\mathbb{R}^n} \rho(\|x\|) \mathrm{d}x = 1,
\]
where the equality with $\rho$ is a consequence of $\tfrac{\rho_1 + \rho_2}{2} = \rho$.
 For $\delta_1 \delta_2 \neq 0$ the $\rho_i$ clearly do not define the same measure, hence $\mu$ is not extreme.\\

\noindent\emph{($\Leftarrow$) If $\rho$ takes exactly two values on $[0,\Delta]$, then $\mu$ is extreme:}
Assume now that $\rho$ takes exactly two values on $[0,\Delta]$. Since $\rho$ is decreasing
and satisfies $\rho(\Delta)=e^{-\varepsilon}\rho(0)$, there exists $r'\in(0,\Delta]$ such that
\[\rho(t)=\rho(0)\,\mathbbm{1}_{[0,r')}(t)+e^{-\varepsilon}\rho(0)\,\mathbbm{1}_{[r',\Delta]}(t),\]
for $t\in[0,\Delta]$, and $\rho$ is extended to $[0,\infty)$ by \eqref{eq:max_decay_extreme}. Suppose $\rho=\tfrac12(\rho_1+\rho_2)$ with $\rho_1$ and $\rho_2$ being two radial functions corresponding to two densities in $\mathcal{D}$.
Without loss of generality, assume $\rho_1(0)\ge \rho(0)$, hence $\rho_2(0)\le \rho(0)$. Since $\rho_2$ is decreasing, for all $t\in[0,r')$ we have $\rho_2(t)\le \rho_2(0)\le \rho(0)$. But on $[0,r')$ we also have $\rho(t)=\rho(0)$, thus
\[\rho_1(t)=2\rho(t)-\rho_2(t)=2\rho(0)-\rho_2(t)\ge \rho(0)=\rho(t),\]
for $t\in[0,r')$.
Since $\rho_1$ is decreasing and satisfies the maximal decay rule
$\rho_1(\Delta)=e^{-\varepsilon}\rho_1(0)\ge e^{-\varepsilon}\rho(0)$, we get
\[\rho_1(t)\ge \rho_1(\Delta)\ge e^{-\varepsilon}\rho(0)=\rho(t), \]
for $t\in[r',\Delta]$.  Thus $\rho_1(t)\ge \rho(t)$ for all $t\in[0,\Delta]$, and by iterating the common decay rule
\eqref{eq:max_decay_extreme} the same inequality holds for all $t\ge 0$.
Since $\rho_1,\rho$ correspond to probability densities, the nonnegative function
$r^{n-1}(\rho_1(r)-\rho(r))$ has integral zero and must vanish a.e. Therefore $\rho_1=\rho$ a.e., and consequently $\rho_2=\rho$ a.e. as well. This shows that $\mu$ cannot be written as a nontrivial convex combination of two distinct elements of $\mathcal{D}$, i.e., $\mu$ is extreme.
\end{proof}

\section{Norm Symmetric Coordinates} \label{sec: Norm Symmetric Coordinates}

Let $\lVert\cdot\rVert$ be a norm on $\mathbb{R}^{n}$ and let $\mathbb{S}_{\lVert\cdot\rVert}^{n-1}$ be the unit sphere with respect to $\lVert\cdot\rVert$, i.e.,
\begin{equation*}
    \mathbb{S}_{\lVert\cdot\rVert}^{n-1} = \{x\in\mathbb{R}^{n} : \lVert x \rVert = 1\}.
\end{equation*}
For brevity we write $\mathbb{S}^{n-1}\coloneqq \mathbb{S}^{n-1}_{\|\cdot\|}$.

We consider the continuous bijection $\Psi$ from $\mathbb{R}^{n}\backslash\{0\}$ onto $(0,\infty)\times\mathbb{S}_{\lVert\cdot\rVert}^{n-1}$ given by
\begin{equation*}
    \Psi(x) = \Big(\lVert x \rVert, \frac{x}{\lVert x \rVert}\Big),
\end{equation*}
whose inverse is $\Psi^{-1}(r,\theta)=r\theta$. Define a measure $\nu$ on $(0,\infty)$ by
\begin{equation*}
    \nu(A) \coloneqq \int_{A} \lvert \mathbb{B} \rvert nr^{n-1} \,\mathrm{d}r 
\end{equation*}
where we recall our notation $\mathbb{B} \coloneqq \{ x \in \mathbb{R}^n : \|x\| < 1 \}$. 
Finally, define the uniform measure $\sigma$ on $\mathbb{S}^{n-1}$ by
\begin{equation*}
    \sigma(B) = \frac{\lvert \Psi^{-1}((0,1) \times B) \rvert}{\lvert \mathbb{B} \rvert} 
\end{equation*}

\begin{proposition}
\label{Prop:NormCoordinates}
Let $m$ be the Lebesgue measure. If $\Psi$, $\nu$ and $\sigma$ are as above, then
\begin{equation*}
    (\Psi^{-1})_{\ast}(\nu\otimes\sigma) = m.
\end{equation*}
Equivalently, if $g \in L_{1}(m)$, then
\begin{align*}
    \int_{\mathbb{R}^{n}} g(x) \,\mathrm{d}x 
        &= \int_{\mathbb{S}^{n-1}} \int_{0}^{\infty} g(r\theta) \lvert \mathbb{B} \rvert nr^{n-1} \,\mathrm{d}r \,\mathrm{d}\sigma(\theta).
        \\
\end{align*}
\end{proposition}

\begin{proof}
Since $\Psi$ is a bijection of $\mathbb{R}^{n}\backslash\{0\}$ onto $(0,\infty)\times\mathbb{S}^{n-1}$, it suffices to prove that $\Psi_{\ast}m = \nu\otimes\sigma$. Furthermore, since $\nu\otimes\sigma$ is a product measure, it suffices to prove that, for all rectangles $A \times B \subset (0,\infty) \times \mathbb{S}^{n-1}$,
\begin{equation*}
    \Psi_{\ast}m(A \times B) = \nu\otimes\sigma(A \times B).
\end{equation*}
Recall that any measure on $(0,\infty)$ is characterized by its behavior on open intervals. Hence, it is enough to show that, for all $a\geq0$,
\begin{equation*}
    \Psi_{\ast}m((0,a) \times B) = \nu((0,a)) \sigma(B).
\end{equation*}

Let $a\geq0$ and $B\subseteq \mathbb{S}^{n-1}$. The definition of the pushforward measure implies that
\begin{equation*}
    \Psi_{\ast}m((0,a) \times B) = \lvert \Psi^{-1}((0,a) \times B) \rvert.
\end{equation*}
Observe that $\Psi^{-1}((0,a) \times B) = a \Psi^{-1}((0,1) \times B)$. Thus, by the homogeneity of the Lebesgue measure,
\begin{align*}
    \Psi_{\ast}m((0,a) \times B) &= a^{n} \lvert \Psi^{-1}((0,1) \times B) \rvert\\
    &= \lvert \mathbb{B}_{1} \rvert a^{n} \sigma(B),
\end{align*}
where the last equality follows from the definition of $\sigma$. A straightforward manipulation leads to
\begin{align*}
    \Psi_{\ast}m((0,a) \times B) &= \int_{(0,a)} \lvert \mathbb{B}_{1} \rvert nr^{n-1} \,\mathrm{d}r \cdot \sigma(B)\\
    &= \nu((0,a)) \sigma(B),
\end{align*}
as required.
\end{proof}
\begin{lemma}\label{lem: norm symmetry is indpendence and uniform}
Let $X\sim \mu$ be a random vector on $\mathbb{R}^n$.
Then $X$ has a radially symmetric density of the form $f(x)=\rho(\|x\|)$ for some $\rho:[0,\infty)\to[0,\infty)$
if and only if $\|X\|$ and $X/\|X\|$ are independent, $X/\|X\|\sim\sigma$, and $\|X\|$ has density
\[
    r \longmapsto |\mathbb{B}|\,n r^{n-1}\rho(r)\qquad\text{on }(0,\infty).
\]
Equivalently, for measurable $E\subseteq\mathbb{S}^{n-1}$,
$$\sigma(E)=\frac{|E_1|}{|\mathbb{B}|},$$
where 
$$E_a \coloneqq \Big\{x\in\mathbb{R}^n:\frac{x}{\|x\|}\in E,\ \|x\|<a\Big\}.$$
\end{lemma}
\begin{proof}
Assume first that $X$ has density $f(x)=\rho(\|x\|)$.
Let $g:(0,\infty)\to\mathbb{R}$ and $h:\mathbb{S}^{n-1}\to\mathbb{R}$ be bounded and measurable. Using Proposition~\ref{Prop:NormCoordinates},
\begin{align*}
\mathbb{E}\!\left[g(\|X\|)\,h\!\Big(\frac{X}{\|X\|}\Big)\right]
&=
\int_{\mathbb{R}^n} g(\|x\|)\,h\!\Big(\frac{x}{\|x\|}\Big)\,\rho(\|x\|)\,\mathrm{d}x\\
&=
\int_{\mathbb{S}^{n-1}}\int_0^\infty g(r)\,h(\theta)\,\rho(r)\,|\mathbb{B}|\,n r^{n-1}\,\mathrm{d}r\,\mathrm{d}\sigma(\theta)\\
&=
\left(\int_0^\infty g(r)\,\rho(r)\,\,\mathrm{d}\nu(r)\right)\left(\int_{\mathbb{S}^{n-1}} h(\theta)\,\mathrm{d}\sigma(\theta)\right),
\end{align*}
which shows that $\|X\|$ and $X/\|X\|$ are independent and $X/\|X\|\sim\sigma$. Moreover, $\|X\|$ has density
$r\mapsto |\mathbb{B}|\,n r^{n-1}\rho(r)$ on $(0,\infty)$.

Conversely, assume $\|X\|$ and $X/\|X\|$ are independent, $X/\|X\|\sim\sigma$, and $\|X\|$ has density
$r\mapsto |\mathbb{B}|\,n r^{n-1}\rho(r)$ on $(0,\infty)$.
Then $\Pr(\|X\|=0)=0$, hence $X/\|X\|$ is well-defined a.s. Let $g:\mathbb{R}^n\to\mathbb{R}$ be bounded and measurable and set $f(x)\coloneqq \rho(\|x\|)$. Using Proposition~\ref{Prop:NormCoordinates} and the product structure of the law of $(\|X\|,X/\|X\|)$,
\begin{align*}
\mathbb{E}[g(X)]
&=
\mathbb{E}\!\left[g\!\Big(\|X\|\frac{X}{\|X\|}\Big)\right]\\
&=
\int_{\mathbb{S}^{n-1}}\int_0^\infty g(r\theta)\,|\mathbb{B}|\,n r^{n-1}\rho(r)\,\mathrm{d}r\,\mathrm{d}\sigma(\theta)\\
&=
\int_{\mathbb{S}^{n-1}}\int_0^\infty g(r\theta)\,f(r\theta)\,|\mathbb{B}|\,n r^{n-1}\,\mathrm{d}r\,\mathrm{d}\sigma(\theta)\\
&=
\int_{\mathbb{R}^n} g(x)\,f(x)\,\mathrm{d}x,
\end{align*}
so $X$ has density $f(x)=\rho(\|x\|)$.
\end{proof}

\end{document}